\documentclass[12pt]{article}%
\usepackage{amsfonts}
\usepackage{amsmath}
\usepackage{amssymb}
\usepackage{graphicx}%
\newenvironment{proof}[1][Proof]{\noindent\textbf{#1.} }{\ \rule{0.5em}{0.5em}}
\begin{document}

\author{{\small V. V. Fern\'{a}ndez}$^{{\footnotesize 1}}${\small , A. M.
Moya}$^{{\footnotesize 1}}${\small , E. Notte-Cuello}$^{{\footnotesize 2}}%
${\small \ and W. A. Rodrigues Jr}$^{{\footnotesize 1}}${\small . }\\$^{{\footnotesize 1}}\hspace{-0.1cm}${\footnotesize Institute of Mathematics,
Statistics and Scientific Computation}\\{\footnotesize IMECC-UNICAMP CP 6065}\\{\footnotesize 13083-859 Campinas, SP, Brazil}\\$^{{\footnotesize 2}}${\small Departamento de} {\small Matem\'{a}ticas,}\\{\small Universidad de La Serena}\\{\small Av. Cisternas 1200, La Serena-Chile}\\{\small e-mail:} {\small walrod@ime.unicamp.br and enotte@userena.cl }}
\title{Duality Products of Multivectors and Multiforms, and Extensors}
\maketitle

\begin{abstract}
In this paper we study in details the properties of the duality product of
multivectors and multiforms (used in the definition of the hyperbolic Clifford
algebra of \textit{multivefors}) and introduce the theory of the
$k$\emph{\ multivector and }$l$\emph{\ multiform variables multivector
}(\emph{or multiform})\emph{\ extensors} over $V$ (defining the spaces
$\left.  \overset{\left.  {}\right.  }{ext}\right.  _{k}^{l}(V)$ and $\left.
\overset{\left.  \ast\right.  }{ext}\right.  _{k}^{l}(V)$) studying their
properties with considerable detail.

\end{abstract}

\newpage

\section{Introduction}

In this article we first briefly recall the exterior algebras of multivectors
(elements of $%
%TCIMACRO{\dbigwedge }%
%BeginExpansion
{\displaystyle\bigwedge}
%EndExpansion
V$) and multiforms (elements of $%
%TCIMACRO{\dbigwedge }%
%BeginExpansion
{\displaystyle\bigwedge}
%EndExpansion
V^{\ast}$) associated with a real vector space $V$ of finite dimension. Next
we explore the structure of the duality product $<$ , $>$ of multivectors by
multiforms used in the definition of the hyperbolic Clifford algebra
$\mathcal{C}\ell(V\oplus V^{\ast},<$ , $>)$ of \textit{multivecfors}
\cite{qr07}. We detail some important properties of the left and right
contracted products among the elements of $%
%TCIMACRO{\dbigwedge }%
%BeginExpansion
{\displaystyle\bigwedge}
%EndExpansion
V$ and $%
%TCIMACRO{\dbigwedge }%
%BeginExpansion
{\displaystyle\bigwedge}
%EndExpansion
V^{\ast}$. Next, we give a theory of the $k$\emph{\ multivector and }%
$l$\emph{\ multiform variables multivector }(\emph{or multiform}%
)\emph{\ extensors} over $V$ (defining the spaces $\left.  \overset{\left.
{}\right.  }{ext}\right.  _{k}^{l}(V)$ and $\left.  \overset{\left.
\ast\right.  }{ext}\right.  _{k}^{l}(V)$) introducing the concept of exterior
product of extensors, and of several operators acting on these objects as,
e.g., the adjoint operator, the extension operator, and the generalized
operator procedure. We study the properties of these operators with
considerable detail. Extensors \textit{fields} (with representatives that are
appropriate mappings from an open set $U\subset M$ to $\left.  \overset
{\left.  {}\right.  }{ext}\right.  _{k}^{l}(V)$ or $\left.  \overset{\left.
\ast\right.  }{ext}\right.  _{k}^{l}(V)$) and the operators acting on them
will be shown\ in forthcoming papers to provide a valuable and simplifying
tool (and an improvement over a previous presentation
\cite{fmr107,fmr207,fmr307,fmr407} which uses only the multivector and
extensor calculus)\ in the study of some of the fundamental ingredients of the
differential geometry\footnote{A presentation of the differential geometry of
an arbitrary manifold (admitting a metric field) using the Clifford bundle
formalism may be found in \cite{rodoliv2006}.} of an arbitrary manifold $M$
equipped with an arbitrary connection.

\section{Multivectors and Multiforms}

Let $V$ be a vector space over $\mathbb{R}$ with finite dimension, i.e., $\dim
V=n$ with $n\in\mathbb{N},$ and let $\ V^{\ast}$ be its dual vector space.
Recall that
\begin{equation}
\dim V=\dim V^{\ast}=n. \label{MM1}%
\end{equation}

Let us consider an integer number $k$ with $0\leq k\leq n$. The real vector
spaces of $k$-vectors over $V,$ i.e., the set of skew-symmetric $k$%
-contravariant tensors over $V$, and the real vector spaces of $k$-forms over
$V$, i.e., the set of skew-symmetric $k$-covariant tensors over $V,$ will be
as usually denoted by $%
%TCIMACRO{\tbigwedge \nolimits^{k}}%
%BeginExpansion
{\textstyle\bigwedge\nolimits^{k}}
%EndExpansion
V$ and $%
%TCIMACRO{\tbigwedge \nolimits^{k}}%
%BeginExpansion
{\textstyle\bigwedge\nolimits^{k}}
%EndExpansion
V^{\ast}$, respectively.

We identify, as usual $0$-vectors to real numbers, i.e., $%
%TCIMACRO{\tbigwedge \nolimits^{0}}%
%BeginExpansion
{\textstyle\bigwedge\nolimits^{0}}
%EndExpansion
V=\mathbb{R}$, and $1$-vectors to objects living in $V$, i.e., $%
%TCIMACRO{\tbigwedge \nolimits^{1}}%
%BeginExpansion
{\textstyle\bigwedge\nolimits^{1}}
%EndExpansion
V\simeq V.$ Also, we identify $0$-forms with real numbers, i.e., $%
%TCIMACRO{\tbigwedge \nolimits^{0}}%
%BeginExpansion
{\textstyle\bigwedge\nolimits^{0}}
%EndExpansion
V=\mathbb{R}$, and $1$-forms wih objects living in $V^{\ast},$ i.e., $%
%TCIMACRO{\tbigwedge \nolimits^{1}}%
%BeginExpansion
{\textstyle\bigwedge\nolimits^{1}}
%EndExpansion
V^{\ast}=V^{\ast}$. Recall that
\begin{equation}
\dim%
%TCIMACRO{\tbigwedge \nolimits^{k}}%
%BeginExpansion
{\textstyle\bigwedge\nolimits^{k}}
%EndExpansion
V=\dim%
%TCIMACRO{\tbigwedge \nolimits^{k}}%
%BeginExpansion
{\textstyle\bigwedge\nolimits^{k}}
%EndExpansion
V^{\ast}=\binom{n}{k}.\label{MM2}%
\end{equation}

The $0$-vectors, $2$-vectors,\ldots, $(n-1)$-vectors and $n$-vectors are
called scalars, bivectors,\ldots, pseudovectors and pseudoscalars,
respectively. The $0$-forms, $2$-forms,\ldots, $(n-1)$-forms and $n$-forms are
called scalars, biforms,\ldots, pseudoforms and pseudoscalars.

\bigskip Given a vector space $V$ over the real field $\mathbb{R}$, we define
$%
%TCIMACRO{\tbigwedge }%
%BeginExpansion
{\textstyle\bigwedge}
%EndExpansion
V$ as the exterior direct sum%

\[%
%TCIMACRO{\tbigwedge }%
%BeginExpansion
{\textstyle\bigwedge}
%EndExpansion
V=%
%TCIMACRO{\dsum \limits_{r=0}^{n}}%
%BeginExpansion
{\displaystyle\sum\limits_{r=0}^{n}}
%EndExpansion
\oplus%
%TCIMACRO{\tbigwedge \nolimits^{r}}%
%BeginExpansion
{\textstyle\bigwedge\nolimits^{r}}
%EndExpansion
V=%
%TCIMACRO{\dbigoplus \nolimits_{r=0}^{n}}%
%BeginExpansion
{\displaystyle\bigoplus\nolimits_{r=0}^{n}}
%EndExpansion%
%TCIMACRO{\tbigwedge \nolimits^{r}}%
%BeginExpansion
{\textstyle\bigwedge\nolimits^{r}}
%EndExpansion
V.
\]

To simplify the notation we sometimes write \textit{simply}
\[%
%TCIMACRO{\tbigwedge }%
%BeginExpansion
{\textstyle\bigwedge}
%EndExpansion
V=\mathbb{R}+V+%
%TCIMACRO{\tbigwedge \nolimits^{2}}%
%BeginExpansion
{\textstyle\bigwedge\nolimits^{2}}
%EndExpansion
V+\cdots+%
%TCIMACRO{\tbigwedge \nolimits^{n-1}}%
%BeginExpansion
{\textstyle\bigwedge\nolimits^{n-1}}
%EndExpansion
V+%
%TCIMACRO{\tbigwedge \nolimits^{n}}%
%BeginExpansion
{\textstyle\bigwedge\nolimits^{n}}
%EndExpansion
V.
\]
As it is well known the set of multivectors over $V$ has a natural structure
of vector space over $\mathbb{R}$ and we have
\begin{align}
\dim%
%TCIMACRO{\tbigwedge }%
%BeginExpansion
{\textstyle\bigwedge}
%EndExpansion
V &  =\dim\mathbb{R+}\dim V+\dim%
%TCIMACRO{\tbigwedge \nolimits^{2}}%
%BeginExpansion
{\textstyle\bigwedge\nolimits^{2}}
%EndExpansion
V+\cdots+\dim%
%TCIMACRO{\tbigwedge \nolimits^{n-1}}%
%BeginExpansion
{\textstyle\bigwedge\nolimits^{n-1}}
%EndExpansion
V+\dim%
%TCIMACRO{\tbigwedge \nolimits^{n}}%
%BeginExpansion
{\textstyle\bigwedge\nolimits^{n}}
%EndExpansion
V\nonumber\\
&  =\binom{n}{0}+\binom{n}{1}+\binom{n}{2}+\cdots+\binom{n}{n-1}+\binom{n}%
{n}=2^{n}.\label{MM3}%
\end{align}

An element of $%
%TCIMACRO{\tbigwedge }%
%BeginExpansion
{\textstyle\bigwedge}
%EndExpansion
V$ will be called a multivector over $V$. If $X\in%
%TCIMACRO{\tbigwedge }%
%BeginExpansion
{\textstyle\bigwedge}
%EndExpansion
V$ we write:%
\[
X=X^{0}+X^{1}+X^{2}+\cdots+X^{n-1}+X^{n}.
\]

In what follows we shall need also the vector space $%
%TCIMACRO{\tbigwedge }%
%BeginExpansion
{\textstyle\bigwedge}
%EndExpansion
V^{\ast}=%
%TCIMACRO{\dbigoplus \nolimits_{r=0}^{n}}%
%BeginExpansion
{\displaystyle\bigoplus\nolimits_{r=0}^{n}}
%EndExpansion%
%TCIMACRO{\tbigwedge \nolimits^{r}}%
%BeginExpansion
{\textstyle\bigwedge\nolimits^{r}}
%EndExpansion
V^{\ast}$. An element of $%
%TCIMACRO{\tbigwedge }%
%BeginExpansion
{\textstyle\bigwedge}
%EndExpansion
V^{\ast}$ will be called a \textit{multiform over} $V$. As in the case of
multivectors we simplify our notation and as in the case of multivectors we
simply write:
\begin{equation}%
%TCIMACRO{\tbigwedge }%
%BeginExpansion
{\textstyle\bigwedge}
%EndExpansion
V^{\ast}=\mathbb{R}+V^{\ast}+%
%TCIMACRO{\tbigwedge \nolimits^{2}}%
%BeginExpansion
{\textstyle\bigwedge\nolimits^{2}}
%EndExpansion
V^{\ast}+\cdots+%
%TCIMACRO{\tbigwedge \nolimits^{n-1}}%
%BeginExpansion
{\textstyle\bigwedge\nolimits^{n-1}}
%EndExpansion
V^{\ast}+%
%TCIMACRO{\tbigwedge \nolimits^{n}}%
%BeginExpansion
{\textstyle\bigwedge\nolimits^{n}}
%EndExpansion
V^{\ast},
\end{equation}
and if $\Phi\in%
%TCIMACRO{\tbigwedge }%
%BeginExpansion
{\textstyle\bigwedge}
%EndExpansion
V^{\ast}$ we write%
\begin{equation}
\Phi=\Phi_{0}+\Phi_{1}+\Phi_{2}+\cdots+\Phi_{n-1}+\Phi_{n}.
\end{equation}
Of course, \ $%
%TCIMACRO{\tbigwedge }%
%BeginExpansion
{\textstyle\bigwedge}
%EndExpansion
V^{\ast}$ has a natural structure of real vector space over $\mathbb{R}$. We
have,
\begin{align}
\dim%
%TCIMACRO{\tbigwedge }%
%BeginExpansion
{\textstyle\bigwedge}
%EndExpansion
V^{\ast} &  =\dim\mathbb{R+}\dim V^{\ast}+\dim%
%TCIMACRO{\tbigwedge \nolimits^{2}}%
%BeginExpansion
{\textstyle\bigwedge\nolimits^{2}}
%EndExpansion
V^{\ast}+\cdots+\dim%
%TCIMACRO{\tbigwedge \nolimits^{n-1}}%
%BeginExpansion
{\textstyle\bigwedge\nolimits^{n-1}}
%EndExpansion
V^{\ast}+\dim%
%TCIMACRO{\tbigwedge \nolimits^{n}}%
%BeginExpansion
{\textstyle\bigwedge\nolimits^{n}}
%EndExpansion
V^{\ast}\nonumber\\
&  =\binom{n}{0}+\binom{n}{1}+\binom{n}{2}+\cdots+\binom{n}{n-1}+\binom{n}%
{n}=2^{n}.\label{MM4}%
\end{align}

We recall that $%
%TCIMACRO{\tbigwedge ^{k}}%
%BeginExpansion
{\textstyle\bigwedge^{k}}
%EndExpansion
V$ is also called the homogeneous multivector space (of degree $k$), and to $%
%TCIMACRO{\tbigwedge ^{k}}%
%BeginExpansion
{\textstyle\bigwedge^{k}}
%EndExpansion
V^{\ast}$ the homogeneous multiform space (of degree $k$).

Let us take an integer number $k$ with $0\leq k\leq n.$ The linear mappings%
\[%
%TCIMACRO{\tbigwedge }%
%BeginExpansion
{\textstyle\bigwedge}
%EndExpansion
V\ni X\longmapsto\left\langle X\right\rangle ^{k}\in%
%TCIMACRO{\tbigwedge }%
%BeginExpansion
{\textstyle\bigwedge}
%EndExpansion
V\text{ and }%
%TCIMACRO{\tbigwedge }%
%BeginExpansion
{\textstyle\bigwedge}
%EndExpansion
V^{\ast}\ni\Phi\longmapsto\left\langle \Phi\right\rangle _{k}\in%
%TCIMACRO{\tbigwedge }%
%BeginExpansion
{\textstyle\bigwedge}
%EndExpansion
V^{\ast}%
\]
such that if $X=X^{0}+X^{1}+\cdots+X^{n}$ and $\Phi=\Phi_{0}+\Phi_{1}%
+\cdots+\Phi_{n},$ then
\begin{equation}
\left\langle X\right\rangle ^{k}=X^{k}\text{ and }\left\langle \Phi
\right\rangle _{k}=\Phi_{k} \label{MM5}%
\end{equation}
are called the $k$\emph{-part operator }(\emph{for multivectors}) and the
$k$\emph{-part operator }(\emph{for multiforms}), respectively. $\left\langle
X\right\rangle ^{k}$ is read as the $k$\emph{-part of }$X$\emph{\ }and
$\left\langle \Phi\right\rangle _{k}$ is read as the $k$\emph{-part of }%
$\Phi.$

It should be evident that for all $X\in%
%TCIMACRO{\tbigwedge }%
%BeginExpansion
{\textstyle\bigwedge}
%EndExpansion
V$ and $\Phi\in%
%TCIMACRO{\tbigwedge }%
%BeginExpansion
{\textstyle\bigwedge}
%EndExpansion
V^{\ast}:$%
\begin{align}
X  &  =\underset{k=0}{\overset{n}{%
%TCIMACRO{\tsum }%
%BeginExpansion
{\textstyle\sum}
%EndExpansion
}}\left\langle X\right\rangle ^{k},\label{MM6}\\
\Phi &  =\underset{k=0}{\overset{n}{%
%TCIMACRO{\tsum }%
%BeginExpansion
{\textstyle\sum}
%EndExpansion
}}\left\langle \Phi\right\rangle _{k}. \label{MM7}%
\end{align}

The linear mappings%
\[%
%TCIMACRO{\tbigwedge }%
%BeginExpansion
{\textstyle\bigwedge}
%EndExpansion
V\ni X\longmapsto\widehat{X}\in%
%TCIMACRO{\tbigwedge }%
%BeginExpansion
{\textstyle\bigwedge}
%EndExpansion
V\text{ and }%
%TCIMACRO{\tbigwedge }%
%BeginExpansion
{\textstyle\bigwedge}
%EndExpansion
V^{\ast}\ni\Phi\longmapsto\widehat{\Phi}\in%
%TCIMACRO{\tbigwedge }%
%BeginExpansion
{\textstyle\bigwedge}
%EndExpansion
V^{\ast}%
\]
such that%
\begin{equation}
\left\langle \widehat{X}\right\rangle ^{k}=\left(  -1\right)  ^{k}\left\langle
X\right\rangle ^{k}\text{ and }\left\langle \widehat{\Phi}\right\rangle
_{k}=\left(  -1\right)  ^{k}\left\langle \Phi\right\rangle _{k} \label{MM8}%
\end{equation}
are called the \emph{grade involution operator (for multivectors)} and the
\emph{grade involution operator (for multiforms)}, respectively.

The linear mappings%
\[%
%TCIMACRO{\tbigwedge }%
%BeginExpansion
{\textstyle\bigwedge}
%EndExpansion
V\ni X\longmapsto\widetilde{X}\in%
%TCIMACRO{\tbigwedge }%
%BeginExpansion
{\textstyle\bigwedge}
%EndExpansion
V\text{ and }%
%TCIMACRO{\tbigwedge }%
%BeginExpansion
{\textstyle\bigwedge}
%EndExpansion
V^{\ast}\ni\Phi\longmapsto\widetilde{\Phi}\in%
%TCIMACRO{\tbigwedge }%
%BeginExpansion
{\textstyle\bigwedge}
%EndExpansion
V^{\ast}%
\]
such that%
\begin{equation}
\left\langle \widetilde{X}\right\rangle ^{k}=\left(  -1\right)  ^{\frac{1}%
{2}k(k-1)}\left\langle X\right\rangle ^{k}\text{ and }\left\langle
\widetilde{\Phi}\right\rangle _{k}=\left(  -1\right)  ^{\frac{1}{2}%
k(k-1)}\left\langle \Phi\right\rangle _{k} \label{MM9}%
\end{equation}
are called the \emph{reversion operator (for multivectors)} and the
\emph{reversion operator (for multiforms)}, respectively.

Both of $%
%TCIMACRO{\tbigwedge }%
%BeginExpansion
{\textstyle\bigwedge}
%EndExpansion
V$ and $%
%TCIMACRO{\tbigwedge }%
%BeginExpansion
{\textstyle\bigwedge}
%EndExpansion
V^{\ast}$ endowed with the exterior product $\wedge$ (of multivectors and
multiforms!) are \emph{associative algebras}, i.e., the \emph{exterior algebra
of multivectors }and the \emph{exterior algebra of multiforms}, respectively.

\section{Duality Scalar Product}

The duality scalar product of a multiform $\Phi$ with a multivector $X$ is the
scalar $\left\langle \Phi,X\right\rangle $ (i.e., the real number) defined by
the following axioms

\begin{itemize}
\item \textbf{ }For all $\alpha,\beta\in\mathbb{R}:$%
\begin{equation}
\left\langle \alpha,\beta\right\rangle =\left\langle \beta,\alpha\right\rangle
=\alpha\beta. \label{DSP1}%
\end{equation}

\item \textbf{ }For all $\Phi_{p}\in%
%TCIMACRO{\tbigwedge \nolimits^{p}}%
%BeginExpansion
{\textstyle\bigwedge\nolimits^{p}}
%EndExpansion
V^{\ast}$ and $X^{p}\in%
%TCIMACRO{\tbigwedge \nolimits^{p}}%
%BeginExpansion
{\textstyle\bigwedge\nolimits^{p}}
%EndExpansion
V$ (with $1\leq p\leq n$)$:$
\begin{equation}
\left\langle \Phi_{p},X^{p}\right\rangle =\left\langle X^{p},\Phi
_{p}\right\rangle =\frac{1}{p!}\Phi_{p}(e_{j_{1}},\ldots,e_{j_{p}}%
)X^{p}(\varepsilon^{j_{1}},\ldots,\varepsilon^{j_{p}}), \label{DSP2}%
\end{equation}
where $\left\{  e_{j},\varepsilon^{j}\right\}  $ is any pair of dual bases
over $V.$

\item For all $\Phi\in%
%TCIMACRO{\tbigwedge }%
%BeginExpansion
{\textstyle\bigwedge}
%EndExpansion
V^{\ast}$ and $X\in%
%TCIMACRO{\tbigwedge }%
%BeginExpansion
{\textstyle\bigwedge}
%EndExpansion
V:$ if $\Phi=\Phi_{0}+\Phi_{1}+\cdots+\Phi_{n}$ and $X=X^{0}+X^{1}%
+\cdots+X^{n},$ then%
\begin{equation}
\left\langle \Phi,X\right\rangle =\left\langle X,\Phi\right\rangle
=\overset{n}{\underset{p=0}{%
%TCIMACRO{\tsum }%
%BeginExpansion
{\textstyle\sum}
%EndExpansion
}}\left\langle \Phi_{p},X^{p}\right\rangle . \label{DSP3}%
\end{equation}

\end{itemize}

We emphasize that the scalar $\Phi_{p}(e_{j_{1}},\ldots,e_{j_{p}}%
)X^{p}(\varepsilon^{j_{1}},\ldots,\varepsilon^{j_{p}})$ has frame independent
character, i.e., it does not depend on the pair of dual bases $\left\{
e_{j},\varepsilon^{j}\right\}  $ used for calculating it, since $\Phi_{p}$ and
$X^{p}$ are $p$-linear mappings.

Note that for all $\omega\in V^{\ast}$ and $v\in V$ it holds
\begin{equation}
\left\langle v,\omega\right\rangle =\left\langle \omega,v\right\rangle
=\omega(v). \label{DSP4}%
\end{equation}

We present two noticeable properties for the duality scalar product between
$p$-forms and $p$-vectors

\begin{itemize}
\item \textbf{ }For all $\Phi_{p}\in%
%TCIMACRO{\tbigwedge ^{p}}%
%BeginExpansion
{\textstyle\bigwedge^{p}}
%EndExpansion
V^{\ast},$ and $v_{1},\ldots,v_{p}\in V:$%
\begin{equation}
\left\langle \Phi_{p},v_{1}\wedge\cdots\wedge v_{p}\right\rangle =\left\langle
v_{1}\wedge\cdots\wedge v_{p},\Phi_{p}\right\rangle =\Phi_{p}(v_{1}%
,\ldots,v_{p}). \label{DSP5}%
\end{equation}

\item \textbf{ }For all $\omega^{1},\ldots,\omega^{p}\in V^{\ast}$ and
$v_{1},\ldots,v_{p}\in V:$%
\begin{equation}
\left\langle \omega^{1}\wedge\cdots\wedge\omega^{p},v_{1}\wedge\cdots\wedge
v_{p}\right\rangle =\det\left(
\begin{array}
[c]{lll}%
\omega^{1}(v_{1}) & \cdots & \omega^{1}(v_{p})\\
\vdots & \vdots & \vdots\\
\omega^{p}(v_{1}) & \cdots & \omega^{p}(v_{p})
\end{array}
\right)  . \label{DSP6}%
\end{equation}

\end{itemize}

The basic properties for the duality scalar product are the non-degeneracy and
the distributive laws on the left and on the right with respect to addition of
either multiforms or multivectors, i.e.,

\begin{itemize}
\item
\begin{align}
\left\langle \Phi,X\right\rangle  &  =0,\text{ for all }\Phi\Longrightarrow
X=0,\nonumber\\
\left\langle \Phi,X\right\rangle  &  =0,\text{ for all }X\Longrightarrow
\Phi=0, \label{DSP7}%
\end{align}

\end{itemize}

and,

\begin{itemize}
\item
\begin{align}
\left\langle \Phi+\Psi,X\right\rangle  &  =\left\langle \Phi,X\right\rangle
+\left\langle \Psi,X\right\rangle ,\nonumber\\
\left\langle \Phi,X+Y\right\rangle  &  =\left\langle \Phi,X\right\rangle
+\left\langle \Phi,Y\right\rangle . \label{DSP8}%
\end{align}

\end{itemize}

\section{Duality Contracted Products}

\subsection{Left Contracted Product}

The left contracted product of a multiform $\Phi$ with a multivector $X$ (or,
a multivector $X$ with\ a multiform $\Phi$) is the multivector $\left\langle
\Phi,X\right\vert $ (respectively, the multiform $\left\langle X,\Phi
\right\vert $) defined by the following axioms:

\begin{itemize}
\item For all $\Phi_{p}\in%
%TCIMACRO{\tbigwedge \nolimits^{p}}%
%BeginExpansion
{\textstyle\bigwedge\nolimits^{p}}
%EndExpansion
V^{\ast}$ and $X^{p}\in%
%TCIMACRO{\tbigwedge \nolimits^{p}}%
%BeginExpansion
{\textstyle\bigwedge\nolimits^{p}}
%EndExpansion
V$ with $0\leq p\leq n:$%
\begin{equation}
\left\langle \Phi_{p},X^{p}\right\vert =\left\langle X^{p},\Phi_{p}\right\vert
=\left\langle \widetilde{\Phi}_{p},X^{p}\right\rangle =\left\langle \Phi
_{p},\widetilde{X}^{p}\right\rangle . \label{DCP1}%
\end{equation}

\item For all $\Phi_{p}\in%
%TCIMACRO{\tbigwedge \nolimits^{p}}%
%BeginExpansion
{\textstyle\bigwedge\nolimits^{p}}
%EndExpansion
V^{\ast}$ and $X^{q}\in%
%TCIMACRO{\tbigwedge \nolimits^{q}}%
%BeginExpansion
{\textstyle\bigwedge\nolimits^{q}}
%EndExpansion
V$ (or $X^{p}\in%
%TCIMACRO{\tbigwedge \nolimits^{p}}%
%BeginExpansion
{\textstyle\bigwedge\nolimits^{p}}
%EndExpansion
V$ and $\Phi_{q}\in%
%TCIMACRO{\tbigwedge \nolimits^{q}}%
%BeginExpansion
{\textstyle\bigwedge\nolimits^{q}}
%EndExpansion
V^{\ast}$) with $0\leq p<q\leq n:$%
\begin{align}
\left\langle \Phi_{p},X^{q}\right\vert  &  =\frac{1}{(q-p)!}\left\langle
\widetilde{\Phi}_{p}\wedge\varepsilon^{j_{1}}\wedge\cdots\wedge\varepsilon
^{j_{q-p}},X^{q}\right\rangle e_{j_{1}}\wedge\cdots\wedge e_{j_{q-p}%
},\label{DCP2}\\
\left\langle X^{p},\Phi_{q}\right\vert  &  =\frac{1}{(q-p)!}\left\langle
\widetilde{X}^{p}\wedge e_{j_{1}}\wedge\cdots\wedge e_{j_{q-p}},\Phi
_{q}\right\rangle \varepsilon^{j_{1}}\wedge\cdots\wedge\varepsilon^{j_{q-p}},
\label{DCP3}%
\end{align}
where $\left\{  e_{j},\varepsilon^{j}\right\}  $ is any pair of dual bases for
$V$ and $V^{\ast}$.

\item \textbf{ }For all $\Phi\in%
%TCIMACRO{\tbigwedge }%
%BeginExpansion
{\textstyle\bigwedge}
%EndExpansion
V^{\ast}$ and $X\in%
%TCIMACRO{\tbigwedge }%
%BeginExpansion
{\textstyle\bigwedge}
%EndExpansion
V:$ if $\Phi=\Phi_{0}+\Phi_{1}+\cdots+\Phi_{n}$ and $X=X^{0}+X^{1}%
+\cdots+X^{n},$ then%
\begin{align}
\left\langle \Phi,X\right\vert  &  =\overset{n}{\underset{k=0}{%
%TCIMACRO{\tsum }%
%BeginExpansion
{\textstyle\sum}
%EndExpansion
}}\underset{j=0}{\overset{n-k}{%
%TCIMACRO{\tsum }%
%BeginExpansion
{\textstyle\sum}
%EndExpansion
}}\left\langle \Phi_{j},X^{k+j}\right\vert ,\label{DCP4}\\
\left\langle X,\Phi\right\vert  &  =\overset{n}{\underset{k=0}{%
%TCIMACRO{\tsum }%
%BeginExpansion
{\textstyle\sum}
%EndExpansion
}}\underset{j=0}{\overset{n-k}{%
%TCIMACRO{\tsum }%
%BeginExpansion
{\textstyle\sum}
%EndExpansion
}}\left\langle X^{j},\Phi_{k+j}\right\vert . \label{DCP5}%
\end{align}

\end{itemize}

Note that the $(q-p)$-vector $\left\langle \Phi_{p},X^{q}\right\vert $ and the
$(q-p)$-form $\left\langle X^{p},\Phi_{q}\right\vert $ have frame independent
character, i.e., they do not depend on the pair of frames $\left\{
e_{j},\varepsilon^{j}\right\}  $ chosen for calculating them.

The left contracted product has the following basic properties:

\begin{itemize}
\item \textbf{ }Let us take $\Phi_{p}\in%
%TCIMACRO{\tbigwedge \nolimits^{p}}%
%BeginExpansion
{\textstyle\bigwedge\nolimits^{p}}
%EndExpansion
V^{\ast}$ and $X^{q}\in%
%TCIMACRO{\tbigwedge \nolimits^{q}}%
%BeginExpansion
{\textstyle\bigwedge\nolimits^{q}}
%EndExpansion
V$ with $0\leq p\leq q\leq n.$\ For all $\Psi_{q-p}\in%
%TCIMACRO{\tbigwedge \nolimits^{q-p}}%
%BeginExpansion
{\textstyle\bigwedge\nolimits^{q-p}}
%EndExpansion
V^{\ast},$ it holds
\begin{equation}
\left\langle \left\langle \Phi_{p},X^{q}\right\vert ,\Psi_{q-p}\right\rangle
=\left\langle X^{q},\widetilde{\Phi}_{p}\wedge\Psi_{q-p}\right\rangle .
\label{DCP6}%
\end{equation}

\item \textbf{ }Let us take $X^{p}\in%
%TCIMACRO{\tbigwedge \nolimits^{p}}%
%BeginExpansion
{\textstyle\bigwedge\nolimits^{p}}
%EndExpansion
V$ and $\Phi_{q}\in%
%TCIMACRO{\tbigwedge \nolimits^{q}}%
%BeginExpansion
{\textstyle\bigwedge\nolimits^{q}}
%EndExpansion
V^{\ast}$ with $0\leq p\leq q\leq n.$ For all $Y^{q-p}\in%
%TCIMACRO{\tbigwedge \nolimits^{q-p}}%
%BeginExpansion
{\textstyle\bigwedge\nolimits^{q-p}}
%EndExpansion
V,$ it holds
\begin{equation}
\left\langle \left\langle X^{p},\Phi_{q}\right\vert ,Y^{q-p}\right\rangle
=\left\langle \Phi_{q},\widetilde{X}^{p}\wedge Y^{q-p}\right\rangle .
\label{DCP7}%
\end{equation}

\item For all $X\in%
%TCIMACRO{\tbigwedge }%
%BeginExpansion
{\textstyle\bigwedge}
%EndExpansion
V$ and $\Phi,\Psi\in%
%TCIMACRO{\tbigwedge }%
%BeginExpansion
{\textstyle\bigwedge}
%EndExpansion
V^{\ast}:$
\begin{equation}
\left\langle \left\langle \Phi,X\right\vert ,\Psi\right\rangle =\left\langle
X,\widetilde{\Phi}\wedge\Psi\right\rangle . \label{DCP8}%
\end{equation}

\item \textbf{ }For all $\Phi\in%
%TCIMACRO{\tbigwedge }%
%BeginExpansion
{\textstyle\bigwedge}
%EndExpansion
V^{\ast}$ and $X,Y\in%
%TCIMACRO{\tbigwedge }%
%BeginExpansion
{\textstyle\bigwedge}
%EndExpansion
V:$
\begin{equation}
\left\langle \left\langle X,\Phi\right\vert ,Y\right\rangle =\left\langle
\Phi,\widetilde{X}\wedge Y\right\rangle . \label{DCP9}%
\end{equation}

\end{itemize}

The left contracted product satisfies the distributive laws on the left and on
the right.

\begin{itemize}
\item \textbf{ }For all $\Phi,\Psi\in%
%TCIMACRO{\tbigwedge }%
%BeginExpansion
{\textstyle\bigwedge}
%EndExpansion
V^{\ast}$ and $X,Y\in%
%TCIMACRO{\tbigwedge }%
%BeginExpansion
{\textstyle\bigwedge}
%EndExpansion
V:$
\begin{align}
\left\langle (\Phi+\Psi),X\right\vert  &  =\left\langle \Phi,X\right\vert
+\left\langle \Psi,X\right\vert ,\nonumber\\
\left\langle \Phi,(X+Y)\right\vert  &  =\left\langle \Phi,X\right\vert
+\left\langle \Phi,Y\right\vert . \label{DCP10}%
\end{align}

\item \textbf{ }For all $X,Y\in%
%TCIMACRO{\tbigwedge }%
%BeginExpansion
{\textstyle\bigwedge}
%EndExpansion
V$ and $\Phi,\Psi\in%
%TCIMACRO{\tbigwedge }%
%BeginExpansion
{\textstyle\bigwedge}
%EndExpansion
V^{\ast}:$
\begin{align}
\left\langle (X+Y),\Phi\right\vert  &  =\left\langle X,\Phi\right\vert
+\left\langle Y,\Phi\right\vert ,\nonumber\\
\left\langle X,(\Phi+\Psi)\right\vert  &  =\left\langle X,\Phi\right\vert
+\left\langle X,\Psi\right\vert . \label{DCP11}%
\end{align}

\end{itemize}

\begin{proof}
We present only the proof of the property given by Eq. (\ref{DCP6}), the other
proofs being somewhat analogous.

First note that if $\Psi_{q-p}\in%
%TCIMACRO{\tbigwedge \nolimits^{q-p}}%
%BeginExpansion
{\textstyle\bigwedge\nolimits^{q-p}}
%EndExpansion
V^{\ast}$ and $\left\{  e_{j},\varepsilon^{j}\right\}  $ is any pair of dual
bases for $V$ and $V^{\ast},$\ we can write
\[
\Psi_{q-p}=\frac{1}{\left(  q-p\right)  !}\left\langle \Psi_{q-p},e_{j_{1}%
}\wedge...\wedge e_{j_{q-p}}\right\rangle \varepsilon^{j_{1}}\wedge
...\wedge\varepsilon^{j_{q-p}}.
\]
Then, using the definition (\ref{DCP2}) and the above equation we have%
\[%
\begin{array}
[c]{l}%
\left\langle \left\langle \Phi_{p},X^{q}\right\vert ,\Psi_{q-p}\right\rangle
\\
=\frac{1}{\left(  q-p\right)  !}\left\langle \left\langle \widetilde{\Phi}%
_{p}\wedge\varepsilon^{j_{1}}\wedge\cdots\wedge\varepsilon^{j_{q-p}}%
,X^{q}\right\rangle e_{j_{1}}\wedge\cdots\wedge e_{j_{q-p}},\Psi
_{q-p}\right\rangle \\
=\frac{1}{\left(  q-p\right)  !}\left\langle \widetilde{\Phi}_{p}%
\wedge\varepsilon^{j_{1}}\wedge\cdots\wedge\varepsilon^{j_{q-p}}%
,X^{q}\right\rangle \left\langle e_{j_{1}}\wedge\cdots\wedge e_{j_{q-p}}%
,\Psi_{q-p}\right\rangle \\
=\left\langle X^{q},\frac{1}{\left(  q-p\right)  !}\left\langle \Psi
_{q-p},e_{j_{1}}\wedge\cdots\wedge e_{j_{q-p}}\right\rangle \widetilde{\Phi
}_{p}\wedge\varepsilon^{j_{1}}\wedge\cdots\wedge\varepsilon^{j_{q-p}%
}\right\rangle \\
=\left\langle X^{q},\widetilde{\Phi}_{p}\wedge\frac{1}{\left(  q-p\right)
!}\left\langle \Psi_{q-p},e_{j_{1}}\wedge\cdots\wedge e_{j_{q-p}}\right\rangle
\varepsilon^{j_{1}}\wedge\cdots\wedge\varepsilon^{j_{q-p}}\right\rangle \\
=\left\langle X^{q},\widetilde{\Phi}_{p}\wedge\Psi_{q-p}\right\rangle ,
\end{array}
\]
and the result is proved.
\end{proof}

\subsection{Right Contracted Product}

The right contracted product of a multiform $\Phi$ with a multivector $X$ (or,
a multivector $X$ with a multiform $\Phi$) is the multiform $\left\vert
\Phi,X\right\rangle $ (respectively, the multivector $\left\vert
X,\Phi\right\rangle $) defined by the following axioms:

\begin{itemize}
\item For $\Phi_{p}\in%
%TCIMACRO{\tbigwedge \nolimits^{p}}%
%BeginExpansion
{\textstyle\bigwedge\nolimits^{p}}
%EndExpansion
V^{\ast}$ and $X^{p}\in%
%TCIMACRO{\tbigwedge \nolimits^{p}}%
%BeginExpansion
{\textstyle\bigwedge\nolimits^{p}}
%EndExpansion
V%
%TCIMACRO{\U{a8}}%
%BeginExpansion
\ddot{}%
%EndExpansion
$ with $n\geq p\geq0:$%
\begin{equation}
\left\vert \Phi_{p},X^{p}\right\rangle =\left\vert X^{p},\Phi_{p}\right\rangle
=\left\langle \widetilde{\Phi}_{p},X^{p}\right\rangle =\left\langle \Phi
_{p},\widetilde{X}^{p}\right\rangle . \label{DCP12}%
\end{equation}

\item For $\Phi_{p}\in%
%TCIMACRO{\tbigwedge \nolimits^{p}}%
%BeginExpansion
{\textstyle\bigwedge\nolimits^{p}}
%EndExpansion
V^{\ast}$ and $X^{q}\in%
%TCIMACRO{\tbigwedge \nolimits^{q}}%
%BeginExpansion
{\textstyle\bigwedge\nolimits^{q}}
%EndExpansion
V%
%TCIMACRO{\U{a8}}%
%BeginExpansion
\ddot{}%
%EndExpansion
$ (or $X^{p}\in%
%TCIMACRO{\tbigwedge \nolimits^{p}}%
%BeginExpansion
{\textstyle\bigwedge\nolimits^{p}}
%EndExpansion
V$ and $\Phi_{q}\in%
%TCIMACRO{\tbigwedge \nolimits^{q}}%
%BeginExpansion
{\textstyle\bigwedge\nolimits^{q}}
%EndExpansion
V^{\ast}$) with $n\geq p>q\geq0:$%
\begin{align}
\left\vert \Phi_{p},X^{q}\right\rangle  &  =\frac{1}{(p-q)!}\left\langle
\Phi_{p},e_{j_{1}}\wedge\cdots\wedge e_{j_{p-q}}\wedge\widetilde{X}%
^{q}\right\rangle \varepsilon^{j_{1}}\wedge\cdots\wedge\varepsilon^{j_{p-q}%
},\label{DCP13}\\
\left\vert X^{p},\Phi_{q}\right\rangle  &  =\frac{1}{(p-q)!}\left\langle
X^{p},\varepsilon^{j_{1}}\wedge\cdots\wedge\varepsilon^{j_{p-q}}%
\wedge\widetilde{\Phi}_{q}\right\rangle e_{j_{1}}\wedge\cdots\wedge
e_{j_{p-q}}, \label{DCP14}%
\end{align}
where $\left\{  e_{j},\varepsilon^{j}\right\}  $ is any pair of dual bases for
$V$ and $V^{\ast}$.

\item \textbf{ }For all $\Phi\in%
%TCIMACRO{\tbigwedge }%
%BeginExpansion
{\textstyle\bigwedge}
%EndExpansion
V^{\ast}$ and $X\in%
%TCIMACRO{\tbigwedge }%
%BeginExpansion
{\textstyle\bigwedge}
%EndExpansion
V:$ if $\Phi=\Phi_{0}+\Phi_{1}+\cdots+\Phi_{n}$ and $X=X^{0}+X^{1}%
+\cdots+X^{n},$ then%
\begin{align}
\left\vert \Phi,X\right\rangle  &  =\overset{n}{\underset{k=0}{%
%TCIMACRO{\tsum }%
%BeginExpansion
{\textstyle\sum}
%EndExpansion
}}\underset{j=0}{\overset{n-k}{%
%TCIMACRO{\tsum }%
%BeginExpansion
{\textstyle\sum}
%EndExpansion
}}\left\vert \Phi_{k+j},X^{j}\right\rangle ,\label{DCP15}\\
\left\vert X,\Phi\right\rangle  &  =\overset{n}{\underset{k=0}{%
%TCIMACRO{\tsum }%
%BeginExpansion
{\textstyle\sum}
%EndExpansion
}}\underset{j=0}{\overset{n-k}{%
%TCIMACRO{\tsum }%
%BeginExpansion
{\textstyle\sum}
%EndExpansion
}}\left\vert X^{k+j},\Phi_{j}\right\rangle . \label{DCP16}%
\end{align}

\end{itemize}

The right contracted product satisfies the following basic properties:

\begin{itemize}
\item \textbf{ }Let us take $\Phi_{p}\in%
%TCIMACRO{\tbigwedge \nolimits^{p}}%
%BeginExpansion
{\textstyle\bigwedge\nolimits^{p}}
%EndExpansion
V^{\ast}$ and $X^{q}\in%
%TCIMACRO{\tbigwedge \nolimits^{q}}%
%BeginExpansion
{\textstyle\bigwedge\nolimits^{q}}
%EndExpansion
V$ with $n\geq p\geq q\geq0.$\ For all $Y^{p-q}\in%
%TCIMACRO{\tbigwedge \nolimits^{p-q}}%
%BeginExpansion
{\textstyle\bigwedge\nolimits^{p-q}}
%EndExpansion
V,$ it holds
\begin{equation}
\left\langle Y^{p-q},\left\vert \Phi_{p},X^{q}\right\rangle \right\rangle
=\left\langle Y^{p-q}\wedge\widetilde{X}^{q},\Phi_{p}\right\rangle .
\label{DCP17}%
\end{equation}

\item \textbf{ }Let us take $X^{p}\in%
%TCIMACRO{\tbigwedge \nolimits^{p}}%
%BeginExpansion
{\textstyle\bigwedge\nolimits^{p}}
%EndExpansion
V$ and $\Phi_{q}\in%
%TCIMACRO{\tbigwedge \nolimits^{q}}%
%BeginExpansion
{\textstyle\bigwedge\nolimits^{q}}
%EndExpansion
V^{\ast}$ with $n\geq p\geq q\geq0.$ For all $\Psi_{p-q}\in%
%TCIMACRO{\tbigwedge \nolimits^{p-q}}%
%BeginExpansion
{\textstyle\bigwedge\nolimits^{p-q}}
%EndExpansion
V^{\ast},$ it holds
\begin{equation}
\left\langle \Psi_{p-q},\left\vert X^{p},\Phi_{q}\right\rangle \right\rangle
=\left\langle \Psi_{p-q}\wedge\widetilde{\Phi}_{q},X^{p}\right\rangle .
\label{DCP18}%
\end{equation}

\item \textbf{ }For all $\Phi\in%
%TCIMACRO{\tbigwedge }%
%BeginExpansion
{\textstyle\bigwedge}
%EndExpansion
V^{\ast}$ and $X,Y\in%
%TCIMACRO{\tbigwedge }%
%BeginExpansion
{\textstyle\bigwedge}
%EndExpansion
V:$%
\begin{equation}
\left\langle Y,\left\vert \Phi,X\right\rangle \right\rangle =\left\langle
Y\wedge\widetilde{X},\Phi\right\rangle . \label{DCP19}%
\end{equation}

\item \textbf{ }For all $X\in%
%TCIMACRO{\tbigwedge }%
%BeginExpansion
{\textstyle\bigwedge}
%EndExpansion
V$ and $\Phi,\Psi\in%
%TCIMACRO{\tbigwedge }%
%BeginExpansion
{\textstyle\bigwedge}
%EndExpansion
V^{\ast}$:
\begin{equation}
\left\langle \Psi,\left\vert X,\Phi\right\rangle \right\rangle =\left\langle
\Psi\wedge\widetilde{\Phi},X\right\rangle . \label{DCP20}%
\end{equation}

\end{itemize}

The right contracted product satisfies also the following distributive laws:

\begin{itemize}
\item \textbf{ }For all $\Phi,\Psi\in%
%TCIMACRO{\tbigwedge }%
%BeginExpansion
{\textstyle\bigwedge}
%EndExpansion
V^{\ast}$ and $X,Y\in%
%TCIMACRO{\tbigwedge }%
%BeginExpansion
{\textstyle\bigwedge}
%EndExpansion
V:$%
\begin{align}
\left\vert (\Phi+\Psi),X\right\rangle  &  =\left\vert \Phi,X\right\rangle
+\left\vert \Psi,X\right\rangle ,\nonumber\\
\left\vert \Phi,(X+Y)\right\rangle  &  =\left\vert \Phi,X\right\rangle
+\left\vert \Phi,Y\right\rangle . \label{CDP21}%
\end{align}

\item \textbf{ }For all $X,Y\in%
%TCIMACRO{\tbigwedge }%
%BeginExpansion
{\textstyle\bigwedge}
%EndExpansion
V$ and $\Phi,\Psi\in%
%TCIMACRO{\tbigwedge }%
%BeginExpansion
{\textstyle\bigwedge}
%EndExpansion
V^{\ast}:$%
\begin{align}
\left\vert (X+Y),\Phi\right\rangle  &  =\left\vert X,\Phi\right\rangle
+\left\vert Y,\Phi\right\rangle ,\nonumber\\
\left\vert X,(\Phi+\Psi)\right\rangle  &  =\left\vert X,\Phi\right\rangle
+\left\vert X,\Psi\right\rangle . \label{DCP22}%
\end{align}

\end{itemize}

We present two noticeable properties relating the left and right contracted products.

\begin{itemize}
\item \textbf{ }For all $\Phi_{p}\in%
%TCIMACRO{\tbigwedge \nolimits^{p}}%
%BeginExpansion
{\textstyle\bigwedge\nolimits^{p}}
%EndExpansion
V^{\ast}$ and $X^{q}\in%
%TCIMACRO{\tbigwedge ^{q}}%
%BeginExpansion
{\textstyle\bigwedge^{q}}
%EndExpansion
V$ with $p\leq q:$%
\begin{equation}
\left\langle \Phi_{p},X^{q}\right\vert =(-1)^{p(q-p)}\left\vert X^{q},\Phi
_{p}\right\rangle . \label{DCP23}%
\end{equation}

\item \textbf{ }For all $X^{p}\in%
%TCIMACRO{\tbigwedge ^{p}}%
%BeginExpansion
{\textstyle\bigwedge^{p}}
%EndExpansion
V$ and $\Phi_{q}\in%
%TCIMACRO{\tbigwedge \nolimits^{q}}%
%BeginExpansion
{\textstyle\bigwedge\nolimits^{q}}
%EndExpansion
V^{\ast}$ with $p\leq q:$%
\begin{equation}
\left\langle X^{p},\Phi_{q}\right\vert =(-1)^{p(q-p)}\left\vert \Phi_{q}%
,X^{p}\right\rangle . \label{DCP24}%
\end{equation}

\end{itemize}

\begin{proof}
We present only the proof of the property given by Eq. (\ref{DCP23}), the
proofs of the other properties are analogous. Let $\Phi_{p}\in%
%TCIMACRO{\tbigwedge \nolimits^{p}}%
%BeginExpansion
{\textstyle\bigwedge\nolimits^{p}}
%EndExpansion
V^{\ast}$ and $X^{q}\in%
%TCIMACRO{\tbigwedge ^{q}}%
%BeginExpansion
{\textstyle\bigwedge^{q}}
%EndExpansion
V$ with $p\leq q,$ then
\[%
\begin{array}
[c]{ll}%
\left\langle \Phi_{p},X^{q}\right\vert  & =\frac{1}{\left(  q-p\right)
!}\left\langle \widetilde{\Phi}_{p}\wedge\varepsilon^{j_{1}}\wedge
...\wedge\varepsilon^{j_{q-p}},X^{q}\right\rangle e_{j_{1}}\wedge...\wedge
e_{j_{q-p}}\\
& =\frac{1}{\left(  q-p\right)  !}\left\langle \left(  -1\right)  ^{p\left(
q-p\right)  }\varepsilon^{j_{1}}\wedge...\wedge\varepsilon^{j_{q-p}}%
\wedge\widetilde{\Phi}_{p},X^{q}\right\rangle e_{j_{1}}\wedge...\wedge
e_{j_{q-p}}\\
& =\left(  -1\right)  ^{p\left(  q-p\right)  }\frac{1}{\left(  q-p\right)
!}\left\langle X^{q},\varepsilon^{j_{1}}\wedge...\wedge\varepsilon^{j_{q-p}%
}\wedge\widetilde{\Phi}_{p}\right\rangle e_{j_{1}}\wedge...\wedge e_{j_{q-p}%
}\\
& =(-1)^{p(q-p)}\left\vert X^{q},\Phi_{p}\right\rangle ,
\end{array}
\]
and the result is proved.
\end{proof}

\section{Extensors}

Let $%
%TCIMACRO{\tbigwedge _{1}^{\diamond}}%
%BeginExpansion
{\textstyle\bigwedge_{1}^{\diamond}}
%EndExpansion
V,\ldots$ and $%
%TCIMACRO{\tbigwedge _{k}^{\diamond}}%
%BeginExpansion
{\textstyle\bigwedge_{k}^{\diamond}}
%EndExpansion
V$ be $k$ subspaces of $%
%TCIMACRO{\tbigwedge }%
%BeginExpansion
{\textstyle\bigwedge}
%EndExpansion
V$ such that each of them is any sum of homogeneous subspaces of $%
%TCIMACRO{\tbigwedge }%
%BeginExpansion
{\textstyle\bigwedge}
%EndExpansion
V,$ \ and let $%
%TCIMACRO{\tbigwedge _{1}^{\diamond}}%
%BeginExpansion
{\textstyle\bigwedge_{1}^{\diamond}}
%EndExpansion
V^{\ast},\ldots$ and $%
%TCIMACRO{\tbigwedge _{l}^{\diamond}}%
%BeginExpansion
{\textstyle\bigwedge_{l}^{\diamond}}
%EndExpansion
V^{\ast}$ be $l$ subspaces of $%
%TCIMACRO{\tbigwedge }%
%BeginExpansion
{\textstyle\bigwedge}
%EndExpansion
V^{\ast}$ such that each of them is any sum of homogeneous subspaces of $%
%TCIMACRO{\tbigwedge }%
%BeginExpansion
{\textstyle\bigwedge}
%EndExpansion
V^{\ast}.$

If $%
%TCIMACRO{\tbigwedge ^{\diamond}}%
%BeginExpansion
{\textstyle\bigwedge^{\diamond}}
%EndExpansion
V$ is any sum of homogeneous subspaces of $%
%TCIMACRO{\tbigwedge }%
%BeginExpansion
{\textstyle\bigwedge}
%EndExpansion
V,$ a multilinear mapping
\begin{align}
\underset{k\text{-copies}}{\underbrace{%
%TCIMACRO{\tbigwedge \nolimits_{1}^{\diamond}}%
%BeginExpansion
{\textstyle\bigwedge\nolimits_{1}^{\diamond}}
%EndExpansion
V\times\cdots\times%
%TCIMACRO{\tbigwedge \nolimits_{k}^{\diamond}}%
%BeginExpansion
{\textstyle\bigwedge\nolimits_{k}^{\diamond}}
%EndExpansion
V}}  &  \times\underset{l\text{-copies}}{\underbrace{%
%TCIMACRO{\tbigwedge \nolimits_{1}^{\diamond}}%
%BeginExpansion
{\textstyle\bigwedge\nolimits_{1}^{\diamond}}
%EndExpansion
V^{\ast}\times\cdots\times%
%TCIMACRO{\tbigwedge \nolimits_{l}^{\diamond}}%
%BeginExpansion
{\textstyle\bigwedge\nolimits_{l}^{\diamond}}
%EndExpansion
V^{\ast}}}\ni(X_{1},\ldots,X_{k},\Phi^{1},\ldots,\Phi^{l})\nonumber\\
&  \longmapsto\tau(X_{1},\ldots,X_{k},\Phi^{1},\ldots,\Phi^{l})\in%
%TCIMACRO{\tbigwedge \nolimits^{\diamond}}%
%BeginExpansion
{\textstyle\bigwedge\nolimits^{\diamond}}
%EndExpansion
V \label{E1}%
\end{align}
is called a $k$\emph{\ multivector and }$l$\emph{\ multiform variables
multivector extensor} over $V.$

If $%
%TCIMACRO{\tbigwedge \nolimits^{\diamond}}%
%BeginExpansion
{\textstyle\bigwedge\nolimits^{\diamond}}
%EndExpansion
V^{\ast}$ is any sum of homogeneous subspaces of $%
%TCIMACRO{\tbigwedge }%
%BeginExpansion
{\textstyle\bigwedge}
%EndExpansion
V^{\ast},$ a multilinear mapping
\begin{align}
\underset{k\text{-copies}}{\underbrace{%
%TCIMACRO{\tbigwedge \nolimits_{1}^{\diamond}}%
%BeginExpansion
{\textstyle\bigwedge\nolimits_{1}^{\diamond}}
%EndExpansion
V\times\cdots\times%
%TCIMACRO{\tbigwedge \nolimits_{k}^{\diamond}}%
%BeginExpansion
{\textstyle\bigwedge\nolimits_{k}^{\diamond}}
%EndExpansion
V}}  &  \times\underset{l\text{-copies}}{\underbrace{%
%TCIMACRO{\tbigwedge \nolimits_{1}^{\diamond}}%
%BeginExpansion
{\textstyle\bigwedge\nolimits_{1}^{\diamond}}
%EndExpansion
V^{\ast}\times\cdots\times%
%TCIMACRO{\tbigwedge \nolimits_{l}^{\diamond}}%
%BeginExpansion
{\textstyle\bigwedge\nolimits_{l}^{\diamond}}
%EndExpansion
V^{\ast}}}\ni(X_{1},\ldots,X_{k},\Phi^{1},\ldots,\Phi^{l})\nonumber\\
&  \longmapsto\tau(X_{1},\ldots,X_{k},\Phi^{1},\ldots,\Phi^{l})\in%
%TCIMACRO{\tbigwedge \nolimits^{\diamond}}%
%BeginExpansion
{\textstyle\bigwedge\nolimits^{\diamond}}
%EndExpansion
V^{\ast} \label{E2}%
\end{align}
is called a $k$\emph{\ multivector and }$l$\emph{\ multiform variables
multiform extensor} over $V.$

The set of all the $k$ multivector and $l$ multiform variables multivector
extensors over $V$ has a natural structure of real vector space, and will be
denoted by the highly suggestive notation $ext(%
%TCIMACRO{\tbigwedge \nolimits_{1}^{\diamond}}%
%BeginExpansion
{\textstyle\bigwedge\nolimits_{1}^{\diamond}}
%EndExpansion
V,\ldots,%
%TCIMACRO{\tbigwedge \nolimits_{k}^{\diamond}}%
%BeginExpansion
{\textstyle\bigwedge\nolimits_{k}^{\diamond}}
%EndExpansion
V,%
%TCIMACRO{\tbigwedge \nolimits_{1}^{\diamond}}%
%BeginExpansion
{\textstyle\bigwedge\nolimits_{1}^{\diamond}}
%EndExpansion
V^{\ast},\ldots,%
%TCIMACRO{\tbigwedge \nolimits_{l}^{\diamond}}%
%BeginExpansion
{\textstyle\bigwedge\nolimits_{l}^{\diamond}}
%EndExpansion
V^{\ast};%
%TCIMACRO{\tbigwedge \nolimits^{\diamond}}%
%BeginExpansion
{\textstyle\bigwedge\nolimits^{\diamond}}
%EndExpansion
V)$. When no confusion arises, we use the more simple notation $\left.
\overset{\left.  {}\right.  }{ext}\right.  _{k}^{l}(V)$ for that space.

We obviously have that:
\begin{equation}
\dim\left.  \overset{\left.  {}\right.  }{ext}\right.  _{k}^{l}(V)=\dim%
%TCIMACRO{\tbigwedge \nolimits_{1}^{\diamond}}%
%BeginExpansion
{\textstyle\bigwedge\nolimits_{1}^{\diamond}}
%EndExpansion
V\ldots\dim%
%TCIMACRO{\tbigwedge \nolimits_{k}^{\diamond}}%
%BeginExpansion
{\textstyle\bigwedge\nolimits_{k}^{\diamond}}
%EndExpansion
V\dim%
%TCIMACRO{\tbigwedge \nolimits_{1}^{\diamond}}%
%BeginExpansion
{\textstyle\bigwedge\nolimits_{1}^{\diamond}}
%EndExpansion
V^{\ast}\ldots\dim%
%TCIMACRO{\tbigwedge \nolimits_{l}^{\diamond}}%
%BeginExpansion
{\textstyle\bigwedge\nolimits_{l}^{\diamond}}
%EndExpansion
V^{\ast}\dim%
%TCIMACRO{\tbigwedge \nolimits^{\diamond}}%
%BeginExpansion
{\textstyle\bigwedge\nolimits^{\diamond}}
%EndExpansion
V. \label{E3}%
\end{equation}

The set of all the $k$ multivector and $l$ multiform variables multiform
extensors over $V$ has also a natural structure of real vector space, and will
be denoted by $ext(%
%TCIMACRO{\tbigwedge \nolimits_{1}^{\diamond}}%
%BeginExpansion
{\textstyle\bigwedge\nolimits_{1}^{\diamond}}
%EndExpansion
V,\ldots,%
%TCIMACRO{\tbigwedge \nolimits_{k}^{\diamond}}%
%BeginExpansion
{\textstyle\bigwedge\nolimits_{k}^{\diamond}}
%EndExpansion
V,%
%TCIMACRO{\tbigwedge \nolimits_{1}^{\diamond}}%
%BeginExpansion
{\textstyle\bigwedge\nolimits_{1}^{\diamond}}
%EndExpansion
V^{\ast},\ldots,%
%TCIMACRO{\tbigwedge \nolimits_{l}^{\diamond}}%
%BeginExpansion
{\textstyle\bigwedge\nolimits_{l}^{\diamond}}
%EndExpansion
V^{\ast};%
%TCIMACRO{\tbigwedge \nolimits^{\diamond}}%
%BeginExpansion
{\textstyle\bigwedge\nolimits^{\diamond}}
%EndExpansion
V^{\ast})$, and when no confusion arises, we use the simple notation $\left.
\overset{\ast}{ext}\right.  _{k}^{l}(V)$ for this space. Also, we have,
\begin{equation}
\dim\left.  \overset{\ast}{ext}\right.  _{k}^{l}(V)=\dim%
%TCIMACRO{\tbigwedge \nolimits_{1}^{\diamond}}%
%BeginExpansion
{\textstyle\bigwedge\nolimits_{1}^{\diamond}}
%EndExpansion
V\ldots\dim%
%TCIMACRO{\tbigwedge \nolimits_{k}^{\diamond}}%
%BeginExpansion
{\textstyle\bigwedge\nolimits_{k}^{\diamond}}
%EndExpansion
V\dim%
%TCIMACRO{\tbigwedge \nolimits_{1}^{\diamond}}%
%BeginExpansion
{\textstyle\bigwedge\nolimits_{1}^{\diamond}}
%EndExpansion
V^{\ast}\ldots\dim%
%TCIMACRO{\tbigwedge \nolimits_{l}^{\diamond}}%
%BeginExpansion
{\textstyle\bigwedge\nolimits_{l}^{\diamond}}
%EndExpansion
V^{\ast}\dim%
%TCIMACRO{\tbigwedge \nolimits^{\diamond}}%
%BeginExpansion
{\textstyle\bigwedge\nolimits^{\diamond}}
%EndExpansion
V^{\ast}. \label{E4}%
\end{equation}

\section{Algebra of Extensors}

\subsection{Exterior Product of Extensors}

We define the exterior product of $\tau\in\left.  \overset{\left.  {}\right.
}{ext}\right.  _{k}^{l}(V)$ and $\sigma\in\left.  \overset{\left.  {}\right.
}{ext}\right.  _{r}^{s}(V)$ (or, $\tau\in\left.  \overset{\ast}{ext}\right.
_{k}^{l}(V)$ and $\sigma\in\left.  \overset{\ast}{ext}\right.  _{r}^{s}(V)$)
as $\tau\wedge\sigma\in\left.  \overset{\left.  {}\right.  }{ext}\right.
_{k+r}^{l+s}(V)$ (respectively, $\tau\wedge\sigma\in\left.  \overset{\ast
}{ext}\right.  _{k+r}^{l+s}(V)$) given by%
\begin{align}
&  \tau\wedge\sigma(X_{1},\ldots,X_{k},Y_{1},\ldots,Y_{r},\Phi^{1},\ldots
,\Phi^{l},\Psi^{1},\ldots,\Psi^{s})\nonumber\\
&  =\tau(X_{1},\ldots,X_{k},\Phi^{1},\ldots,\Phi^{l})\wedge\sigma(Y_{1}%
,\ldots,Y_{r},\Psi^{1},\ldots,\Psi^{s}). \label{E5}%
\end{align}
Note that on the right side appears an exterior product of multivectors
(respectively, an exterior product of multiforms).

The duality scalar product of a multiform extensor $\tau\in\left.
\overset{\ast}{ext}\right.  _{k}^{l}(V)$ with a multivector extensor
$\sigma\in\left.  \overset{\left.  {}\right.  }{ext}\right.  _{r}^{s}(V)$ is
the \textit{scalar} extensor $\left\langle \tau,\sigma\right\rangle \in\left.
\overset{\ast}{ext}\right.  _{k+r}^{l+s}(V)$ defined by%
\begin{align}
&  \left\langle \tau,\sigma\right\rangle (X_{1},\ldots,X_{k},Y_{1}%
,\ldots,Y_{r},\Phi^{1},\ldots,\Phi^{l},\Psi^{1},\ldots,\Psi^{s})\nonumber\\
&  =\left\langle \tau(X_{1},\ldots,X_{k},\Phi^{1},\ldots,\Phi^{l}%
),\sigma(Y_{1},\ldots,Y_{r},\Psi^{1},\ldots,\Psi^{s})\right\rangle .
\label{E6}%
\end{align}

The duality left contracted product of a multiform extensor $\tau\in\left.
\overset{\ast}{ext}\right.  _{k}^{l}(V)$ with a multivector extensor
$\sigma\in\left.  \overset{\left.  {}\right.  }{ext}\right.  _{r}^{s}(V)$ (or,
a multivector extensor $\tau\in\left.  \overset{\left.  {}\right.  }%
{ext}\right.  _{k}^{l}(V)$ with a multiform extensor $\sigma\in\left.
\overset{\ast}{ext}\right.  _{r}^{s}(V)$) is the multivector extensor
$\left\langle \tau,\sigma\right\vert \in\left.  \overset{\left.  {}\right.
}{ext}\right.  _{k+r}^{l+s}(V)$ (respectively, the multiform extensor
$\left\langle \tau,\sigma\right\vert \in\left.  \overset{\ast}{ext}\right.
_{k+r}^{l+s}(V)$) defined by%
\begin{align}
&  \left\langle \tau,\sigma\right\vert (X_{1},\ldots,X_{k},Y_{1},\ldots
,Y_{r},\Phi^{1},\ldots,\Phi^{l},\Psi^{1},\ldots,\Psi^{s})\nonumber\\
&  =\left\langle \tau(X_{1},\ldots,X_{k},\Phi^{1},\ldots,\Phi^{l}%
),\sigma(Y_{1},\ldots,Y_{r},\Psi^{1},\ldots,\Psi^{s})\right\vert . \label{E7}%
\end{align}

The duality right contracted product of a multiform extensor $\tau\in\left.
\overset{\ast}{ext}\right.  _{k}^{l}(V)$ with a multivector extensor
$\sigma\in\left.  \overset{\left.  {}\right.  }{ext}\right.  _{r}^{s}(V)$ (or,
a multivector extensor $\tau\in\left.  \overset{\left.  {}\right.  }%
{ext}\right.  _{k}^{l}(V)$ with a multiform extensor $\sigma\in\left.
\overset{\ast}{ext}\right.  _{r}^{s}(V)$) is the multiform extensor
$\left\vert \tau,\sigma\right\rangle \in\left.  \overset{\ast}{ext}\right.
_{k+r}^{l+s}(V)$ (respectively, the multivector extensor $\left\vert
\tau,\sigma\right\rangle \in\left.  \overset{\left.  {}\right.  }{ext}\right.
_{k+r}^{l+s}(V)$) defined by%
\begin{align}
&  \left\vert \tau,\sigma\right\rangle (X_{1},\ldots,X_{k},Y_{1},\ldots
,Y_{r},\Phi^{1},\ldots,\Phi^{l},\Psi^{1},\ldots,\Psi^{s})\nonumber\\
&  =\left\vert \tau(X_{1},\ldots,X_{k},\Phi^{1},\ldots,\Phi^{l}),\sigma
(Y_{1},\ldots,Y_{r},\Psi^{1},\ldots,\Psi^{s})\right\rangle . \label{E8}%
\end{align}

\subsection{Duality Adjoint of Extensors}

Let $%
%TCIMACRO{\tbigwedge \nolimits^{\diamond}}%
%BeginExpansion
{\textstyle\bigwedge\nolimits^{\diamond}}
%EndExpansion
V$ be any sum of homogeneous subspaces of $%
%TCIMACRO{\tbigwedge }%
%BeginExpansion
{\textstyle\bigwedge}
%EndExpansion
V.$ There exist $\mu$ integer numbers $p_{1},\ldots,p_{\mu}$ with $0\leq
p_{1}<\cdots<p_{\mu}\leq n$ such that $%
%TCIMACRO{\tbigwedge \nolimits^{\diamond}}%
%BeginExpansion
{\textstyle\bigwedge\nolimits^{\diamond}}
%EndExpansion
V=%
%TCIMACRO{\tbigwedge \nolimits^{p_{1}}}%
%BeginExpansion
{\textstyle\bigwedge\nolimits^{p_{1}}}
%EndExpansion
V+\cdots+%
%TCIMACRO{\tbigwedge \nolimits^{p_{\mu}}}%
%BeginExpansion
{\textstyle\bigwedge\nolimits^{p_{\mu}}}
%EndExpansion
V.$ Analogously, if $%
%TCIMACRO{\tbigwedge \nolimits^{\diamond}}%
%BeginExpansion
{\textstyle\bigwedge\nolimits^{\diamond}}
%EndExpansion
V^{\ast}$ is any sum of homogeneous subspaces of $%
%TCIMACRO{\tbigwedge }%
%BeginExpansion
{\textstyle\bigwedge}
%EndExpansion
V^{\ast},$ then there exist $\nu$ integer numbers $q_{1},\ldots,q_{\nu}$ with
$0\leq q_{1}<\cdots<q_{\nu}\leq n$ such that $%
%TCIMACRO{\tbigwedge \nolimits^{\diamond}}%
%BeginExpansion
{\textstyle\bigwedge\nolimits^{\diamond}}
%EndExpansion
V^{\ast}=%
%TCIMACRO{\tbigwedge \nolimits^{q_{1}}}%
%BeginExpansion
{\textstyle\bigwedge\nolimits^{q_{1}}}
%EndExpansion
V^{\ast}+\cdots+%
%TCIMACRO{\tbigwedge \nolimits^{q_{\nu}}}%
%BeginExpansion
{\textstyle\bigwedge\nolimits^{q_{\nu}}}
%EndExpansion
V^{\ast}.$

The linear mappings
\[%
%TCIMACRO{\tbigwedge }%
%BeginExpansion
{\textstyle\bigwedge}
%EndExpansion
V\ni X\longmapsto\left\langle X\right\rangle ^{%
%TCIMACRO{\tbigwedge \nolimits^{\diamond}}%
%BeginExpansion
{\textstyle\bigwedge\nolimits^{\diamond}}
%EndExpansion
V}\in%
%TCIMACRO{\tbigwedge }%
%BeginExpansion
{\textstyle\bigwedge}
%EndExpansion
V\text{ and }%
%TCIMACRO{\tbigwedge }%
%BeginExpansion
{\textstyle\bigwedge}
%EndExpansion
V^{\ast}\ni\Phi\longmapsto\left\langle \Phi\right\rangle _{%
%TCIMACRO{\tbigwedge \nolimits^{\diamond}}%
%BeginExpansion
{\textstyle\bigwedge\nolimits^{\diamond}}
%EndExpansion
V^{\ast}}\in%
%TCIMACRO{\tbigwedge }%
%BeginExpansion
{\textstyle\bigwedge}
%EndExpansion
V^{\ast}%
\]
such that if $%
%TCIMACRO{\tbigwedge \nolimits^{\diamond}}%
%BeginExpansion
{\textstyle\bigwedge\nolimits^{\diamond}}
%EndExpansion
V=%
%TCIMACRO{\tbigwedge \nolimits^{p_{1}}}%
%BeginExpansion
{\textstyle\bigwedge\nolimits^{p_{1}}}
%EndExpansion
V+\cdots+%
%TCIMACRO{\tbigwedge \nolimits^{p_{\mu}}}%
%BeginExpansion
{\textstyle\bigwedge\nolimits^{p_{\mu}}}
%EndExpansion
V$ and $%
%TCIMACRO{\tbigwedge \nolimits^{\diamond}}%
%BeginExpansion
{\textstyle\bigwedge\nolimits^{\diamond}}
%EndExpansion
V^{\ast}=%
%TCIMACRO{\tbigwedge \nolimits^{q_{1}}}%
%BeginExpansion
{\textstyle\bigwedge\nolimits^{q_{1}}}
%EndExpansion
V^{\ast}+\cdots+%
%TCIMACRO{\tbigwedge \nolimits^{q_{\nu}}}%
%BeginExpansion
{\textstyle\bigwedge\nolimits^{q_{\nu}}}
%EndExpansion
V^{\ast},$ then
\begin{equation}
\left\langle X\right\rangle ^{%
%TCIMACRO{\tbigwedge \nolimits^{\diamond}}%
%BeginExpansion
{\textstyle\bigwedge\nolimits^{\diamond}}
%EndExpansion
V}=\left\langle X\right\rangle ^{p_{1}}+\cdots+\left\langle X\right\rangle
^{p_{\mu}}\text{ and }\left\langle \Phi\right\rangle _{%
%TCIMACRO{\tbigwedge \nolimits^{\diamond}}%
%BeginExpansion
{\textstyle\bigwedge\nolimits^{\diamond}}
%EndExpansion
V^{\ast}}=\left\langle \Phi\right\rangle _{q_{1}}+\cdots+\left\langle
\Phi\right\rangle _{q_{\nu}} \label{DAE1}%
\end{equation}
are called the \emph{\ }$%
%TCIMACRO{\tbigwedge \nolimits^{\diamond}}%
%BeginExpansion
{\textstyle\bigwedge\nolimits^{\diamond}}
%EndExpansion
V$\emph{-part operator} \emph{(for multivectors)} and $%
%TCIMACRO{\tbigwedge \nolimits^{\diamond}}%
%BeginExpansion
{\textstyle\bigwedge\nolimits^{\diamond}}
%EndExpansion
V^{\ast}$\emph{-part operator (for multiforms),} respectively.

It should be evident that for all $X\in%
%TCIMACRO{\tbigwedge }%
%BeginExpansion
{\textstyle\bigwedge}
%EndExpansion
V$ and $\Phi\in%
%TCIMACRO{\tbigwedge }%
%BeginExpansion
{\textstyle\bigwedge}
%EndExpansion
V^{\ast}:$%
\begin{align}
\left\langle X\right\rangle ^{%
%TCIMACRO{\tbigwedge \nolimits^{k}}%
%BeginExpansion
{\textstyle\bigwedge\nolimits^{k}}
%EndExpansion
V}  &  =\left\langle X\right\rangle ^{k},\label{DAE2}\\
\left\langle \Phi\right\rangle _{%
%TCIMACRO{\tbigwedge \nolimits^{k}}%
%BeginExpansion
{\textstyle\bigwedge\nolimits^{k}}
%EndExpansion
V^{\ast}}  &  =\left\langle \Phi\right\rangle _{k}. \label{DAE3}%
\end{align}
Thus, $%
%TCIMACRO{\tbigwedge \nolimits^{\diamond}}%
%BeginExpansion
{\textstyle\bigwedge\nolimits^{\diamond}}
%EndExpansion
V$-part operator and $%
%TCIMACRO{\tbigwedge \nolimits^{\diamond}}%
%BeginExpansion
{\textstyle\bigwedge\nolimits^{\diamond}}
%EndExpansion
V^{\ast}$-part operator are the generalizations of $\left\langle \left.
{}\right.  \right\rangle ^{k}$ and $\left\langle \left.  {}\right.
\right\rangle _{k}.$

Let $\tau$ be a multivector extensor of either one multivector variable or one
multiform variable.

If $\tau\in ext(%
%TCIMACRO{\tbigwedge \nolimits_{1}^{\diamond}}%
%BeginExpansion
{\textstyle\bigwedge\nolimits_{1}^{\diamond}}
%EndExpansion
V;%
%TCIMACRO{\tbigwedge \nolimits_{2}^{\diamond}}%
%BeginExpansion
{\textstyle\bigwedge\nolimits_{2}^{\diamond}}
%EndExpansion
V)$ (or, $\tau\in ext(%
%TCIMACRO{\tbigwedge \nolimits_{3}^{\diamond}}%
%BeginExpansion
{\textstyle\bigwedge\nolimits_{3}^{\diamond}}
%EndExpansion
V^{\ast};%
%TCIMACRO{\tbigwedge \nolimits_{4}^{\diamond}}%
%BeginExpansion
{\textstyle\bigwedge\nolimits_{4}^{\diamond}}
%EndExpansion
V)$), then $\tau^{\bigtriangleup}\in ext(%
%TCIMACRO{\tbigwedge \nolimits_{2}^{\diamond}}%
%BeginExpansion
{\textstyle\bigwedge\nolimits_{2}^{\diamond}}
%EndExpansion
V^{\ast};$ $%
%TCIMACRO{\tbigwedge \nolimits_{1}^{\diamond}}%
%BeginExpansion
{\textstyle\bigwedge\nolimits_{1}^{\diamond}}
%EndExpansion
V^{\ast})$ (respectively, $\tau^{\bigtriangleup}\in ext(%
%TCIMACRO{\tbigwedge \nolimits_{4}^{\diamond}}%
%BeginExpansion
{\textstyle\bigwedge\nolimits_{4}^{\diamond}}
%EndExpansion
V^{\ast};%
%TCIMACRO{\tbigwedge \nolimits_{3}^{\diamond}}%
%BeginExpansion
{\textstyle\bigwedge\nolimits_{3}^{\diamond}}
%EndExpansion
V)$) defined by
\begin{align}
\tau^{\bigtriangleup}(\Phi)  &  =\left\langle \Phi,\tau(\left\langle
1\right\rangle ^{%
%TCIMACRO{\tbigwedge \nolimits_{1}^{\diamond}}%
%BeginExpansion
{\textstyle\bigwedge\nolimits_{1}^{\diamond}}
%EndExpansion
V})\right\rangle \nonumber\\
&  +\overset{n}{\underset{k=1}{%
%TCIMACRO{\tsum }%
%BeginExpansion
{\textstyle\sum}
%EndExpansion
}}\frac{1}{k!}\left\langle \Phi,\tau(\left\langle e_{j_{1}}\wedge\cdots\wedge
e_{j_{k}}\right\rangle ^{%
%TCIMACRO{\tbigwedge \nolimits_{1}^{\diamond}}%
%BeginExpansion
{\textstyle\bigwedge\nolimits_{1}^{\diamond}}
%EndExpansion
V})\right\rangle \varepsilon^{j_{1}}\wedge\cdots\wedge\varepsilon^{j_{k}}
\label{DAE4}%
\end{align}
for each $\Phi\in%
%TCIMACRO{\tbigwedge \nolimits_{2}^{\diamond}}%
%BeginExpansion
{\textstyle\bigwedge\nolimits_{2}^{\diamond}}
%EndExpansion
V^{\ast}$ (respectively, for each $\Phi\in%
%TCIMACRO{\tbigwedge \nolimits_{4}^{\diamond}}%
%BeginExpansion
{\textstyle\bigwedge\nolimits_{4}^{\diamond}}
%EndExpansion
V^{\ast}$)%

\begin{align}
\tau^{\bigtriangleup}(\Phi)  &  =\left\langle \Phi,\tau(\left\langle
1\right\rangle _{%
%TCIMACRO{\tbigwedge \nolimits_{3}^{\diamond}}%
%BeginExpansion
{\textstyle\bigwedge\nolimits_{3}^{\diamond}}
%EndExpansion
V^{\ast}})\right\rangle \nonumber\\
&  +\overset{n}{\underset{k=1}{%
%TCIMACRO{\tsum }%
%BeginExpansion
{\textstyle\sum}
%EndExpansion
}}\frac{1}{k!}\left\langle \Phi,\tau(\left\langle \varepsilon^{j_{1}}%
\wedge\cdots\wedge\varepsilon^{j_{k}}\right\rangle _{%
%TCIMACRO{\tbigwedge \nolimits_{3}^{\diamond}}%
%BeginExpansion
{\textstyle\bigwedge\nolimits_{3}^{\diamond}}
%EndExpansion
V^{\ast}})\right\rangle e_{j_{1}}\wedge\cdots\wedge e_{j_{k}} \label{DAE5}%
\end{align}
is called the \emph{duality adjoint of }$\tau.$

The basic properties of the adjoint of multivector extensors are:

\begin{itemize}
\item Let us take $\tau\in ext(%
%TCIMACRO{\tbigwedge \nolimits_{1}^{\diamond}}%
%BeginExpansion
{\textstyle\bigwedge\nolimits_{1}^{\diamond}}
%EndExpansion
V;%
%TCIMACRO{\tbigwedge \nolimits_{2}^{\diamond}}%
%BeginExpansion
{\textstyle\bigwedge\nolimits_{2}^{\diamond}}
%EndExpansion
V).$ For all $X\in%
%TCIMACRO{\tbigwedge \nolimits_{1}^{\diamond}}%
%BeginExpansion
{\textstyle\bigwedge\nolimits_{1}^{\diamond}}
%EndExpansion
V$ and $\Phi\in%
%TCIMACRO{\tbigwedge \nolimits_{2}^{\diamond}}%
%BeginExpansion
{\textstyle\bigwedge\nolimits_{2}^{\diamond}}
%EndExpansion
V^{\ast},$ it holds%
\begin{equation}
\left\langle \tau(X),\Phi\right\rangle =\left\langle X,\tau^{\bigtriangleup
}(\Phi)\right\rangle . \label{DAE6}%
\end{equation}

\item Let us take $\tau\in ext(%
%TCIMACRO{\tbigwedge \nolimits_{3}^{\diamond}}%
%BeginExpansion
{\textstyle\bigwedge\nolimits_{3}^{\diamond}}
%EndExpansion
V^{\ast};%
%TCIMACRO{\tbigwedge \nolimits_{4}^{\diamond}}%
%BeginExpansion
{\textstyle\bigwedge\nolimits_{4}^{\diamond}}
%EndExpansion
V).$ For all $\Phi\in%
%TCIMACRO{\tbigwedge \nolimits_{3}^{\diamond}}%
%BeginExpansion
{\textstyle\bigwedge\nolimits_{3}^{\diamond}}
%EndExpansion
V^{\ast}$ and $\Psi\in%
%TCIMACRO{\tbigwedge \nolimits_{4}^{\diamond}}%
%BeginExpansion
{\textstyle\bigwedge\nolimits_{4}^{\diamond}}
%EndExpansion
V^{\ast},$ it holds
\begin{equation}
\left\langle \tau(\Phi),\Psi\right\rangle =\left\langle \Phi,\tau
^{\bigtriangleup}(\Psi)\right\rangle . \label{DAE7}%
\end{equation}

\end{itemize}

\begin{proof}
We present only the proof of the property given by Eq. (\ref{DAE6}), the other
is analogous. First, observe that if $X\in%
%TCIMACRO{\tbigwedge \nolimits_{1}^{\diamond}}%
%BeginExpansion
{\textstyle\bigwedge\nolimits_{1}^{\diamond}}
%EndExpansion
V,$ then there exists $\mu$ integer numbers $p_{1},\ldots,p_{\mu}$ with $0\leq
p_{1}<\cdots<p_{\mu}\leq n$ such that $X=X^{p_{1}}+...+X^{p_{\mu}}$ with
$X^{p_{i}}\in%
%TCIMACRO{\tbigwedge \nolimits^{p_{i}}}%
%BeginExpansion
{\textstyle\bigwedge\nolimits^{p_{i}}}
%EndExpansion
V$, where the $%
%TCIMACRO{\tbigwedge \nolimits^{p_{i}}}%
%BeginExpansion
{\textstyle\bigwedge\nolimits^{p_{i}}}
%EndExpansion
V$ are homogeneous subspace of $%
%TCIMACRO{\tbigwedge }%
%BeginExpansion
{\textstyle\bigwedge}
%EndExpansion
V,$ thus if $\tau\in ext(%
%TCIMACRO{\tbigwedge \nolimits_{1}^{\diamond}}%
%BeginExpansion
{\textstyle\bigwedge\nolimits_{1}^{\diamond}}
%EndExpansion
V;%
%TCIMACRO{\tbigwedge \nolimits_{2}^{\diamond}}%
%BeginExpansion
{\textstyle\bigwedge\nolimits_{2}^{\diamond}}
%EndExpansion
V)$, we have that%
\[
\tau:%
%TCIMACRO{\tbigwedge \nolimits_{1}^{\diamond}}%
%BeginExpansion
{\textstyle\bigwedge\nolimits_{1}^{\diamond}}
%EndExpansion
V\rightarrow%
%TCIMACRO{\tbigwedge \nolimits_{2}^{\diamond}}%
%BeginExpansion
{\textstyle\bigwedge\nolimits_{2}^{\diamond}}
%EndExpansion
V\qquad\text{or\qquad}\tau:%
%TCIMACRO{\tbigwedge \nolimits^{p_{1}}}%
%BeginExpansion
{\textstyle\bigwedge\nolimits^{p_{1}}}
%EndExpansion
V+...+%
%TCIMACRO{\tbigwedge \nolimits^{p_{\mu}}}%
%BeginExpansion
{\textstyle\bigwedge\nolimits^{p_{\mu}}}
%EndExpansion
V\rightarrow%
%TCIMACRO{\tbigwedge \nolimits^{q_{1}}}%
%BeginExpansion
{\textstyle\bigwedge\nolimits^{q_{1}}}
%EndExpansion
V+...+%
%TCIMACRO{\tbigwedge \nolimits^{q_{\mu}}}%
%BeginExpansion
{\textstyle\bigwedge\nolimits^{q_{\mu}}}
%EndExpansion
V,
\]
where $%
%TCIMACRO{\tbigwedge \nolimits_{1}^{\diamond}}%
%BeginExpansion
{\textstyle\bigwedge\nolimits_{1}^{\diamond}}
%EndExpansion
V=%
%TCIMACRO{\tbigwedge \nolimits^{p_{1}}}%
%BeginExpansion
{\textstyle\bigwedge\nolimits^{p_{1}}}
%EndExpansion
V+...+%
%TCIMACRO{\tbigwedge \nolimits^{p_{\mu}}}%
%BeginExpansion
{\textstyle\bigwedge\nolimits^{p_{\mu}}}
%EndExpansion
V$ and $%
%TCIMACRO{\tbigwedge \nolimits_{2}^{\diamond}}%
%BeginExpansion
{\textstyle\bigwedge\nolimits_{2}^{\diamond}}
%EndExpansion
V=%
%TCIMACRO{\tbigwedge \nolimits^{q_{1}}}%
%BeginExpansion
{\textstyle\bigwedge\nolimits^{q_{1}}}
%EndExpansion
V+...+%
%TCIMACRO{\tbigwedge \nolimits^{q_{\mu}}}%
%BeginExpansion
{\textstyle\bigwedge\nolimits^{q_{\mu}}}
%EndExpansion
V$,\ such that
\[
\tau\left(  X\right)  =\tau\left(  X^{p_{1}}+...+X^{p_{\mu}}\right)
=\tau\left(  X^{p_{1}}\right)  +...+\tau\left(  X^{p_{\mu}}\right)  \in%
%TCIMACRO{\tbigwedge \nolimits^{q_{1}}}%
%BeginExpansion
{\textstyle\bigwedge\nolimits^{q_{1}}}
%EndExpansion
V+...+%
%TCIMACRO{\tbigwedge \nolimits^{q_{\mu}}}%
%BeginExpansion
{\textstyle\bigwedge\nolimits^{q_{\mu}}}
%EndExpansion
V,
\]
thus, we can put
\[
\tau\left(  X^{p_{i}}\right)  \in%
%TCIMACRO{\tbigwedge \nolimits^{q_{i}}}%
%BeginExpansion
{\textstyle\bigwedge\nolimits^{q_{i}}}
%EndExpansion
V\qquad\text{or\qquad}\tau\left\vert _{%
%TCIMACRO{\tbigwedge \nolimits^{p_{i}}}%
%BeginExpansion
{\textstyle\bigwedge\nolimits^{p_{i}}}
%EndExpansion
V}\right.  \equiv\tau_{p_{i}}\in ext(%
%TCIMACRO{\tbigwedge \nolimits^{p_{i}}}%
%BeginExpansion
{\textstyle\bigwedge\nolimits^{p_{i}}}
%EndExpansion
V;%
%TCIMACRO{\tbigwedge \nolimits^{q_{i}}}%
%BeginExpansion
{\textstyle\bigwedge\nolimits^{q_{i}}}
%EndExpansion
V).
\]
Now, with the above observation, we have
\begin{equation}%
\begin{array}
[c]{ll}%
\left\langle \tau(X),\Phi\right\rangle  & =\left\langle \tau\left(  X^{p_{1}%
}+...+X^{p_{\mu}}\right)  ,\Phi_{q_{1}}+...+\Phi_{q_{\mu}}\right\rangle
\medskip\\
& =\left\langle \tau_{p_{1}}\left(  X^{p_{1}}\right)  +...+\tau_{p_{\mu}%
}\left(  X^{p_{\mu}}\right)  ,\Phi_{q_{1}}+...+\Phi_{q_{\mu}}\right\rangle
\medskip\\
& =%
%TCIMACRO{\tsum \limits_{i,j}}%
%BeginExpansion
{\textstyle\sum\limits_{i,j}}
%EndExpansion
\left\langle \tau_{p_{i}}(X^{p_{i}}),\Phi^{q_{j}}\right\rangle ,
\end{array}
\label{EN1}%
\end{equation}
but, by the definition (\ref{DSP3}) we have that%
\[
\left\langle \tau_{p_{i}}(X^{p_{i}}),\Phi^{q_{j}}\right\rangle \left\{
\begin{array}
[c]{l}%
=0\qquad\text{if \ }p_{i}\neq q_{j}\\
\neq0\qquad\text{if \ }p_{i}=q_{j}=s_{l}%
\end{array}
\right.  ,
\]
and from Eq. (\ref{EN1}), we can write%
\begin{equation}
\left\langle \tau(X),\Phi\right\rangle =\left\langle \tau_{s_{1}}(X^{s_{1}%
}),\Phi^{s_{1}}\right\rangle +...+\left\langle \tau_{s_{\mu}}(X^{s_{\mu}%
}),\Phi^{s_{\mu}}\right\rangle .\label{EN2}%
\end{equation}
Now, if we see $\left\langle \tau_{s_{l}}(X^{s_{l}}),\Phi^{s_{l}}\right\rangle
$ as a scalar product of $s_{l}$-vectors, then we have%
\begin{equation}
\left\langle \tau_{s_{l}}(X^{s_{l}}),\Phi^{s_{l}}\right\rangle =\left\langle
X^{s_{l}},\tau_{s_{l}}^{\bigtriangleup}\Phi^{s_{l}}\right\rangle ,\label{EN3}%
\end{equation}
and from Eqs. (\ref{EN2}), (\ref{EN3}), and taking into account Eq.
(\ref{EN1}) we have%
\[%
\begin{array}
[c]{ll}%
\left\langle \tau(X),\Phi\right\rangle  & =\left\langle \tau_{s_{1}}(X^{s_{1}%
}),\Phi^{s_{1}}\right\rangle +...+\left\langle \tau_{s_{\mu}}(X^{s_{\mu}%
}),\Phi^{s_{\mu}}\right\rangle \\
& =\left\langle X^{s_{1}},\tau_{s_{1}}^{\bigtriangleup}\Phi^{s_{1}%
}\right\rangle +...+\left\langle X^{s_{\mu}},\tau_{s_{\mu}}^{\bigtriangleup
}\Phi^{s_{\mu}}\right\rangle \\
& =\left\langle X^{s_{1}}+...+X^{s_{\mu}},\tau_{s_{1}}^{\bigtriangleup}%
\Phi^{s_{1}}+...+\tau_{s_{\mu}}^{\bigtriangleup}\Phi^{s_{\mu}}\right\rangle \\
& =\left\langle X,\tau^{\bigtriangleup}\Phi\right\rangle ,
\end{array}
\]
and the result is proved.
\end{proof}

Let $\sigma$ be a multiform extensor of either one multivector variable or one
multiform variable. Then, if $\sigma\in ext(%
%TCIMACRO{\tbigwedge \nolimits_{1}^{\diamond}}%
%BeginExpansion
{\textstyle\bigwedge\nolimits_{1}^{\diamond}}
%EndExpansion
V;%
%TCIMACRO{\tbigwedge \nolimits_{2}^{\diamond}}%
%BeginExpansion
{\textstyle\bigwedge\nolimits_{2}^{\diamond}}
%EndExpansion
V^{\ast})$ (or, $\sigma\in ext(%
%TCIMACRO{\tbigwedge \nolimits_{3}^{\diamond}}%
%BeginExpansion
{\textstyle\bigwedge\nolimits_{3}^{\diamond}}
%EndExpansion
V^{\ast};%
%TCIMACRO{\tbigwedge \nolimits_{4}^{\diamond}}%
%BeginExpansion
{\textstyle\bigwedge\nolimits_{4}^{\diamond}}
%EndExpansion
V^{\ast})$), then $\sigma^{\bigtriangleup}\in ext(%
%TCIMACRO{\tbigwedge \nolimits_{2}^{\diamond}}%
%BeginExpansion
{\textstyle\bigwedge\nolimits_{2}^{\diamond}}
%EndExpansion
V;$ $%
%TCIMACRO{\tbigwedge \nolimits_{1}^{\diamond}}%
%BeginExpansion
{\textstyle\bigwedge\nolimits_{1}^{\diamond}}
%EndExpansion
V^{\ast})$ (respectively, $\sigma^{\bigtriangleup}\in ext(%
%TCIMACRO{\tbigwedge \nolimits_{4}^{\diamond}}%
%BeginExpansion
{\textstyle\bigwedge\nolimits_{4}^{\diamond}}
%EndExpansion
V;%
%TCIMACRO{\tbigwedge \nolimits_{3}^{\diamond}}%
%BeginExpansion
{\textstyle\bigwedge\nolimits_{3}^{\diamond}}
%EndExpansion
V)$) defined by%
\begin{align}
\sigma^{\bigtriangleup}(X)  &  =\left\langle X,\sigma(\left\langle
1\right\rangle ^{%
%TCIMACRO{\tbigwedge \nolimits_{1}^{\diamond}}%
%BeginExpansion
{\textstyle\bigwedge\nolimits_{1}^{\diamond}}
%EndExpansion
V})\right\rangle \nonumber\\
&  +\overset{n}{\underset{k=1}{%
%TCIMACRO{\tsum }%
%BeginExpansion
{\textstyle\sum}
%EndExpansion
}}\frac{1}{k!}\left\langle X,\sigma(\left\langle e_{j_{1}}\wedge\cdots\wedge
e_{j_{k}}\right\rangle ^{%
%TCIMACRO{\tbigwedge \nolimits_{1}^{\diamond}}%
%BeginExpansion
{\textstyle\bigwedge\nolimits_{1}^{\diamond}}
%EndExpansion
V})\right\rangle \varepsilon^{j_{1}}\wedge\cdots\wedge\varepsilon^{j_{k}}
\label{DAE8}%
\end{align}
for each $X\in%
%TCIMACRO{\tbigwedge \nolimits_{2}^{\diamond}}%
%BeginExpansion
{\textstyle\bigwedge\nolimits_{2}^{\diamond}}
%EndExpansion
V$ (respectively,
\begin{align}
\sigma^{\bigtriangleup}(X)  &  =\left\langle X,\sigma(\left\langle
1\right\rangle _{%
%TCIMACRO{\tbigwedge \nolimits_{3}^{\diamond}}%
%BeginExpansion
{\textstyle\bigwedge\nolimits_{3}^{\diamond}}
%EndExpansion
V^{\ast}})\right\rangle \nonumber\\
&  +\overset{n}{\underset{k=1}{%
%TCIMACRO{\tsum }%
%BeginExpansion
{\textstyle\sum}
%EndExpansion
}}\frac{1}{k!}\left\langle X,\sigma(\left\langle \varepsilon^{j_{1}}%
\wedge\cdots\wedge\varepsilon^{j_{k}}\right\rangle _{%
%TCIMACRO{\tbigwedge \nolimits_{3}^{\diamond}}%
%BeginExpansion
{\textstyle\bigwedge\nolimits_{3}^{\diamond}}
%EndExpansion
V^{\ast}})\right\rangle e_{j_{1}}\wedge\cdots\wedge e_{j_{k}} \label{DAE9}%
\end{align}
for each $X\in%
%TCIMACRO{\tbigwedge \nolimits_{4}^{\diamond}}%
%BeginExpansion
{\textstyle\bigwedge\nolimits_{4}^{\diamond}}
%EndExpansion
V$) is called the \emph{duality adjoint of }$\sigma.$

The basic properties for the adjoint of multiform extensors are:

\begin{itemize}
\item Let us take $\sigma\in ext(%
%TCIMACRO{\tbigwedge \nolimits_{1}^{\diamond}}%
%BeginExpansion
{\textstyle\bigwedge\nolimits_{1}^{\diamond}}
%EndExpansion
V;%
%TCIMACRO{\tbigwedge \nolimits_{2}^{\diamond}}%
%BeginExpansion
{\textstyle\bigwedge\nolimits_{2}^{\diamond}}
%EndExpansion
V^{\ast}).$ For all $X\in%
%TCIMACRO{\tbigwedge \nolimits_{1}^{\diamond}}%
%BeginExpansion
{\textstyle\bigwedge\nolimits_{1}^{\diamond}}
%EndExpansion
V$ and $Y\in%
%TCIMACRO{\tbigwedge \nolimits_{2}^{\diamond}}%
%BeginExpansion
{\textstyle\bigwedge\nolimits_{2}^{\diamond}}
%EndExpansion
V,$ it holds
\begin{equation}
\left\langle \sigma(X),Y\right\rangle =\left\langle X,\sigma^{\bigtriangleup
}(Y)\right\rangle . \label{DAE10}%
\end{equation}

\item \textbf{ } Let us take $\sigma\in ext(%
%TCIMACRO{\tbigwedge \nolimits_{3}^{\diamond}}%
%BeginExpansion
{\textstyle\bigwedge\nolimits_{3}^{\diamond}}
%EndExpansion
V^{\ast};%
%TCIMACRO{\tbigwedge \nolimits_{4}^{\diamond}}%
%BeginExpansion
{\textstyle\bigwedge\nolimits_{4}^{\diamond}}
%EndExpansion
V^{\ast}).$ For all $\Phi\in%
%TCIMACRO{\tbigwedge \nolimits_{3}^{\diamond}}%
%BeginExpansion
{\textstyle\bigwedge\nolimits_{3}^{\diamond}}
%EndExpansion
V^{\ast}$ and $X\in%
%TCIMACRO{\tbigwedge \nolimits_{4}^{\diamond}}%
%BeginExpansion
{\textstyle\bigwedge\nolimits_{4}^{\diamond}}
%EndExpansion
V,$ it holds
\begin{equation}
\left\langle \sigma(\Phi),X\right\rangle =\left\langle \Phi,\sigma
^{\bigtriangleup}(X)\right\rangle . \label{DAE11}%
\end{equation}

\end{itemize}

The linear mapping $\left(  \left.  {}\right.  \right)  ^{\bigtriangleup}$
will be called the \emph{duality adjoint operator.}

\section{Extension Procedure for Operators}

Let $\lambda$ be a linear operator on $V,$ i.e., a linear map $V\ni
v\longmapsto\lambda(v)\in V.$ It can be extended in such a way as to give a
linear operator on $%
%TCIMACRO{\tbigwedge }%
%BeginExpansion
{\textstyle\bigwedge}
%EndExpansion
V,$ namely$\underline{\text{ }\lambda},$ which is defined by%
\[%
%TCIMACRO{\tbigwedge }%
%BeginExpansion
{\textstyle\bigwedge}
%EndExpansion
V\ni X\longmapsto\underline{\lambda}(X)\in%
%TCIMACRO{\tbigwedge }%
%BeginExpansion
{\textstyle\bigwedge}
%EndExpansion
V
\]
such that%
\begin{equation}
\underline{\lambda}(X)=\left\langle 1,X\right\rangle +\underset{k=1}%
{\overset{n}{%
%TCIMACRO{\tsum }%
%BeginExpansion
{\textstyle\sum}
%EndExpansion
}}\frac{1}{k!}\left\langle \varepsilon^{j_{1}}\wedge\cdots\wedge
\varepsilon^{j_{k}},X\right\rangle \lambda(e_{j_{1}})\wedge\cdots\wedge
\lambda(e_{j_{k}}), \label{EPO1}%
\end{equation}
where $\left\{  e_{j},\varepsilon^{j}\right\}  $ is any pair of dual bases for
$V$ and $V^{\ast}$.

We emphasize that $\underline{\lambda}$ is a well-defined linear operator on $%
%TCIMACRO{\tbigwedge }%
%BeginExpansion
{\textstyle\bigwedge}
%EndExpansion
V.$ Note that each $k$-vector $\left\langle \varepsilon^{j_{1}}\wedge
\cdots\wedge\varepsilon^{j_{k}},X\right\rangle \lambda(e_{j_{1}})\wedge
\cdots\wedge\lambda(e_{j_{k}})$ with $1\leq k\leq n$ does not depend on the
choice of $\left\{  e_{j},\varepsilon^{j}\right\}  ,$ and the linearity
follows just from the linearity of the duality scalar product. We call
$\underline{\lambda}$ the \emph{extended }of $\lambda$ (to multivector operator).

The extended of a vector operator $\lambda$ has the following basic properties:

\begin{itemize}
\item \textbf{ }$\underline{\lambda}$ is grade-preserving, i.e.,
\begin{equation}
\text{if }X\in%
%TCIMACRO{\tbigwedge \nolimits^{k}}%
%BeginExpansion
{\textstyle\bigwedge\nolimits^{k}}
%EndExpansion
V,\text{ then }\underline{\lambda}(X)\in%
%TCIMACRO{\tbigwedge \nolimits^{k}}%
%BeginExpansion
{\textstyle\bigwedge\nolimits^{k}}
%EndExpansion
V. \label{EPO2}%
\end{equation}

\item \textbf{ }For all $\alpha\in\mathbb{R},$ $v\in V,$ and $X,Y\in%
%TCIMACRO{\tbigwedge }%
%BeginExpansion
{\textstyle\bigwedge}
%EndExpansion
V:$%
\begin{align}
\underline{\lambda}(\alpha)  &  =\alpha,\label{EPO3}\\
\underline{\lambda}(v)  &  =\lambda(v),\label{EPO4}\\
\underline{\lambda}(X\wedge Y)  &  =\underline{\lambda}(X)\wedge
\underline{\lambda}(Y). \label{EPO5}%
\end{align}

\end{itemize}

We observe that the four basic properties as given by Eq.(\ref{EPO2}),
Eq.(\ref{EPO3}), Eq.(\ref{EPO4}) and Eq.(\ref{EPO5}) are completely equivalent
to the extension procedure of a vector operator.

Let $\lambda$ be a linear operator on $V^{\ast},$ i.e., a linear map $V^{\ast
}\ni\omega\longmapsto\lambda(\omega)\in V^{\ast}.$ It is possible to extend
$\lambda$ in such a way to get a linear operator on $%
%TCIMACRO{\tbigwedge }%
%BeginExpansion
{\textstyle\bigwedge}
%EndExpansion
V^{\ast},$ namely the operator $\underline{\lambda\text{,}}$ defined by%
\[%
%TCIMACRO{\tbigwedge }%
%BeginExpansion
{\textstyle\bigwedge}
%EndExpansion
V^{\ast}\ni\Phi\longmapsto\underline{\lambda}(\Phi)\in%
%TCIMACRO{\tbigwedge }%
%BeginExpansion
{\textstyle\bigwedge}
%EndExpansion
V^{\ast},
\]
such that%
\begin{equation}
\underline{\lambda}(\Phi)=\left\langle 1,\Phi\right\rangle +\underset
{k=1}{\overset{n}{%
%TCIMACRO{\tsum }%
%BeginExpansion
{\textstyle\sum}
%EndExpansion
}}\frac{1}{k!}\left\langle e_{j_{1}}\wedge\cdots\wedge e_{j_{k}}%
,\Phi\right\rangle \lambda(\varepsilon^{j_{1}})\wedge\cdots\wedge
\lambda(\varepsilon^{j_{k}}), \label{EPO6}%
\end{equation}
where $\left\{  e_{j},\varepsilon^{j}\right\}  $ is any pair of dual bases for
$V$ and $V^{\ast}$.

We emphasize that $\underline{\lambda}$ is a well-defined linear operator on $%
%TCIMACRO{\tbigwedge }%
%BeginExpansion
{\textstyle\bigwedge}
%EndExpansion
V^{\ast}.$ We call $\underline{\lambda}$ the \emph{extended }of $\lambda$ (to multiforms).

The extended of a form operator $\lambda$ has the following basic properties.

\begin{itemize}
\item \textbf{ } $\underline{\lambda}$ is grade-preserving, i.e.,
\begin{equation}
\text{if }\Phi\in%
%TCIMACRO{\tbigwedge \nolimits^{k}}%
%BeginExpansion
{\textstyle\bigwedge\nolimits^{k}}
%EndExpansion
V^{\ast},\text{ then }\underline{\lambda}(\Phi)\in%
%TCIMACRO{\tbigwedge \nolimits^{k}}%
%BeginExpansion
{\textstyle\bigwedge\nolimits^{k}}
%EndExpansion
V^{\ast}. \label{EPO7}%
\end{equation}

\item For all $\alpha\in\mathbb{R},$ $\omega\in V^{\ast},$ and $\Phi,\Psi\in%
%TCIMACRO{\tbigwedge }%
%BeginExpansion
{\textstyle\bigwedge}
%EndExpansion
V^{\ast}:$
\begin{align}
\underline{\lambda}(\alpha)  &  =\alpha,\label{EPO8}\\
\underline{\lambda}(\omega)  &  =\lambda(\omega),\label{EPO9}\\
\underline{\lambda}(\Phi\wedge\Psi)  &  =\underline{\lambda}(\Phi
)\wedge\underline{\lambda}(\Psi). \label{EPO10}%
\end{align}

\end{itemize}

The four basic properties given by Eq.(\ref{EPO7}), Eq.(\ref{EPO8})
Eq.(\ref{EPO9}) and Eq.(\ref{EPO10}) are logically equivalent to the extension
procedure of a form operator.

There exists a relationship between the extension procedure of a vector
operator and the extension procedure of a form operator.

Let us take a vector operator (or, a form operator) $\lambda$. As we can see,
the duality adjoint of $\lambda$ is just a form operator (respectively, a
vector operator), and the duality adjoint of $\underline{\lambda}$ is just a
multiform operator (respectively, a multivector operator). It holds that the
duality adjoint of the extended of $\lambda$ is equal to the extended of the
duality adjoint of $\lambda,$ i.e.,%
\begin{equation}
\left(  \underline{\lambda}\right)  ^{\bigtriangleup}=\underline{\left(
\lambda^{\bigtriangleup}\right)  }. \label{EPO11}%
\end{equation}

Thus, it is possible to use the more simple notation $\underline{\lambda
}^{\bigtriangleup}$ to mean either $\left(  \underline{\lambda}\right)
^{\bigtriangleup}$ or $\underline{\left(  \lambda^{\bigtriangleup}\right)  }.$

We present some properties for the extended of an invertible vector operator
$\lambda.$

\begin{itemize}
\item For all $\Phi\in%
%TCIMACRO{\tbigwedge }%
%BeginExpansion
{\textstyle\bigwedge}
%EndExpansion
V^{\ast},$ and $X\in%
%TCIMACRO{\tbigwedge }%
%BeginExpansion
{\textstyle\bigwedge}
%EndExpansion
V:$%
\begin{align}
\underline{\lambda}\left\langle \Phi,X\right\rangle  &  =\left\langle
\underline{\lambda}^{-\bigtriangleup}(\Phi),\underline{\lambda}%
(X)\right\rangle ,\label{EPO12}\\
\underline{\lambda}\left\langle \Phi,X\right\vert  &  =\left\langle
\underline{\lambda}^{-\bigtriangleup}(\Phi),\underline{\lambda}(X)\right\vert
,\label{EPO13}\\
\underline{\lambda}\left\vert X,\Phi\right\rangle  &  =\left\vert
\underline{\lambda}(X),\underline{\lambda}^{-\bigtriangleup}(\Phi
)\right\rangle . \label{EPO14}%
\end{align}

\end{itemize}

We present only the proof for the property given by Eq.(\ref{EPO13}), the
other proofs are analogous.

\begin{proof}
Let us take $X\in%
%TCIMACRO{\tbigwedge }%
%BeginExpansion
{\textstyle\bigwedge}
%EndExpansion
V$ and $\Phi,\Psi\in%
%TCIMACRO{\tbigwedge }%
%BeginExpansion
{\textstyle\bigwedge}
%EndExpansion
V^{\ast}.$ A straightforward calculation, using Eq.(\ref{DAE6}),
Eq.(\ref{DCP8}), Eq.(\ref{EPO10}) and Eq.(\ref{EPO7}), yields
\begin{align*}
\left\langle \underline{\lambda}\left\langle \Phi,X\right\vert ,\Psi
\right\rangle  &  =\left\langle \left\langle \Phi,X\right\vert ,\underline
{\lambda}^{\bigtriangleup}(\Psi)\right\rangle =\left\langle X,\widetilde{\Phi
}\wedge\underline{\lambda}^{\bigtriangleup}(\Psi)\right\rangle \\
&  =\left\langle X,\underline{\lambda}^{\bigtriangleup}(\underline{\lambda
}^{-\bigtriangleup}(\widetilde{\Phi})\wedge\Psi)\right\rangle =\left\langle
\underline{\lambda}(X),\widetilde{\underline{\lambda}^{-\bigtriangleup}(\Phi
)}\wedge\Psi)\right\rangle \\
&  =\left\langle \left\langle \underline{\lambda}^{-\bigtriangleup}%
(\Phi),\underline{\lambda}(X)\right\vert ,\Psi\right\rangle ,
\end{align*}
whence, by the non-degeneracy of duality scalar product the result follows.
\end{proof}

We present now some properties for the extended of an invertible form operator
$\lambda$.

\begin{itemize}
\item For all $\Phi\in%
%TCIMACRO{\tbigwedge }%
%BeginExpansion
{\textstyle\bigwedge}
%EndExpansion
V^{\ast},$ and $X\in%
%TCIMACRO{\tbigwedge }%
%BeginExpansion
{\textstyle\bigwedge}
%EndExpansion
V:$%
\begin{align}
\underline{\lambda}\left\langle \Phi,X\right\rangle  &  =\left\langle
\underline{\lambda}(\Phi),\underline{\lambda}^{-\bigtriangleup}%
(X)\right\rangle ,\label{EPO15}\\
\underline{\lambda}\left\langle \Phi,X\right\vert  &  =\left\langle
\underline{\lambda}(\Phi),\underline{\lambda}^{-\bigtriangleup}(X)\right\vert
,\label{EPO16}\\
\underline{\lambda}\left\vert X,\Phi\right\rangle  &  =\left\vert
\underline{\lambda}^{-\bigtriangleup}(X),\underline{\lambda}(\Phi
)\right\rangle . \label{EPO17}%
\end{align}

\end{itemize}

\subsection{Action of Extended Operators on Extensors}

Let $\lambda$ be an invertible linear operator on $V$. As we saw above, the
extended of $\lambda,$ denoted $\underline{\lambda},$ maps multivectors over
$V$ into multivectors over $V.$ However, it is possible to extend the action
of $\underline{\lambda}$ in such a way that $\underline{\lambda}$ maps
multivector extensors over $V$ into multivector extensors over $V$. We define
the linear mapping%
\[
\left.  \overset{\left.  {}\right.  }{ext}\right.  _{k}^{l}(V)\ni
\tau\longmapsto\underline{\lambda}\tau\in\left.  \overset{\left.  {}\right.
}{ext}\right.  _{k}^{l}(V)
\]
such that%
\begin{equation}
\underline{\lambda}\tau(X_{1},\ldots,X_{k},\Phi^{1},\ldots,\Phi^{l}%
)=\underline{\lambda}\circ\tau(\underline{\lambda}^{-1}(X_{1}),\ldots
,\underline{\lambda}^{-1}(X_{k}),\underline{\lambda}^{\bigtriangleup}(\Phi
^{1}),\ldots,\underline{\lambda}^{\bigtriangleup}(\Phi^{l})) \label{EPO18}%
\end{equation}
for each $X_{1},\ldots,X_{k}\in%
%TCIMACRO{\tbigwedge }%
%BeginExpansion
{\textstyle\bigwedge}
%EndExpansion
V$ and $\Phi^{1},\ldots,\Phi^{l}\in%
%TCIMACRO{\tbigwedge }%
%BeginExpansion
{\textstyle\bigwedge}
%EndExpansion
V^{\ast}.$

It that way $\underline{\lambda}$ can be thought as a linear multivector
extensor operator.

For instance, for $\tau\in\left.  \overset{\left.  {}\right.  }{ext}\right.
_{1}^{0}(V)$ the above definition means%
\begin{equation}
\underline{\lambda}\tau=\underline{\lambda}\circ\tau\circ\underline{\lambda
}^{-1}, \label{EPO19}%
\end{equation}
and for $\tau\in\left.  \overset{\left.  {}\right.  }{ext}\right.  _{0}%
^{1}(V)$ it yields%
\begin{equation}
\underline{\lambda}\tau=\underline{\lambda}\circ\tau\circ\underline{\lambda
}^{\bigtriangleup}. \label{EPO20}%
\end{equation}

Let $\lambda$ be an invertible linear operator on $V^{\ast}.$ Analogously to
the above case, it is possible to extend the action of $\underline{\lambda}$
in such a way to get a linear operator on $\left.  \overset{\left.
\ast\right.  }{ext}\right.  _{k}^{l}(V).$ We define%
\[
\left.  \overset{\left.  \ast\right.  }{ext}\right.  _{k}^{l}(V)\ni
\tau\longmapsto\underline{\lambda}\tau\in\left.  \overset{\left.  \ast\right.
}{ext}\right.  _{k}^{l}(V)
\]
such that%
\begin{equation}
\underline{\lambda}\tau(X_{1},\ldots,X_{k},\Phi^{1},\ldots,\Phi^{l}%
)=\underline{\lambda}\circ\tau(\underline{\lambda}^{\bigtriangleup}%
(X_{1}),\ldots,\underline{\lambda}^{\bigtriangleup}(X_{k}),\underline{\lambda
}^{-1}(\Phi^{1}),\ldots,\underline{\lambda}^{-1}(\Phi^{l})) \label{EPO25}%
\end{equation}
for each $X_{1},\ldots,X_{k}\in%
%TCIMACRO{\tbigwedge }%
%BeginExpansion
{\textstyle\bigwedge}
%EndExpansion
V$ and $\Phi^{1},\ldots,\Phi^{l}\in%
%TCIMACRO{\tbigwedge }%
%BeginExpansion
{\textstyle\bigwedge}
%EndExpansion
V^{\ast}.$

For instance, for $\tau\in\left.  \overset{\left.  {}\right.  }{ext}\right.
_{1}^{0}(V)$ the above definition yields%
\begin{equation}
\underline{\lambda}\tau=\underline{\lambda}\circ\tau\circ\underline{\lambda
}^{\bigtriangleup}, \label{EPO26}%
\end{equation}
and for $\tau\in\left.  \overset{\left.  {}\right.  }{ext}\right.  _{0}%
^{1}(V)$ it implies that%
\begin{equation}
\underline{\lambda}\tau=\underline{\lambda}\circ\tau\circ\underline{\lambda
}^{-1}. \label{EPO27}%
\end{equation}

We give some of the properties of the action of the extended of a vector
operator $\lambda$ on multivector extensors.

\begin{itemize}
\item For all $\tau\in\left.  \overset{\left.  {}\right.  }{ext}\right.
_{k}^{l}(V)$ and $\sigma\in\left.  \overset{\left.  {}\right.  }{ext}\right.
_{r}^{s}(V):$%
\begin{equation}
\underline{\lambda}(\tau\wedge\sigma)=(\underline{\lambda}\tau)\wedge
(\underline{\lambda}\sigma). \label{EPO21}%
\end{equation}

\item \textbf{ } For all $\tau\in\left.  \overset{\ast}{ext}\right.  _{k}%
^{l}(V)$ and $\sigma\in\left.  \overset{\left.  {}\right.  }{ext}\right.
_{r}^{s}(V):$%
\begin{align}
\underline{\lambda}\left\langle \tau,\sigma\right\rangle  &  =\left\langle
\underline{\lambda}^{-\bigtriangleup}\tau,\underline{\lambda}\sigma
\right\rangle ,\label{EPO22}\\
\underline{\lambda}\left\langle \tau,\sigma\right\vert  &  =\left\langle
\underline{\lambda}^{-\bigtriangleup}\tau,\underline{\lambda}\sigma\right\vert
,\label{EPO23}\\
\underline{\lambda}\left\vert \sigma,\tau\right\rangle  &  =\left\vert
\underline{\lambda}\sigma,\underline{\lambda}^{-\bigtriangleup}\tau
\right\rangle . \label{EPO24}%
\end{align}

\end{itemize}

We present the proof of the property given by Eq.(\ref{EPO23}), the other
proofs are analogous.

\begin{proof}
Without any loss of generality, we give the proof for the particular case
where $\tau\in ext(%
%TCIMACRO{\tbigwedge \nolimits_{1}^{\diamond}}%
%BeginExpansion
{\textstyle\bigwedge\nolimits_{1}^{\diamond}}
%EndExpansion
V,%
%TCIMACRO{\tbigwedge \nolimits_{2}^{\diamond}}%
%BeginExpansion
{\textstyle\bigwedge\nolimits_{2}^{\diamond}}
%EndExpansion
V^{\ast};%
%TCIMACRO{\tbigwedge \nolimits^{\diamond}}%
%BeginExpansion
{\textstyle\bigwedge\nolimits^{\diamond}}
%EndExpansion
V^{\ast})$ and $\sigma\in ext(%
%TCIMACRO{\tbigwedge \nolimits_{3}^{\diamond}}%
%BeginExpansion
{\textstyle\bigwedge\nolimits_{3}^{\diamond}}
%EndExpansion
V,%
%TCIMACRO{\tbigwedge \nolimits_{4}^{\diamond}}%
%BeginExpansion
{\textstyle\bigwedge\nolimits_{4}^{\diamond}}
%EndExpansion
V^{\ast};%
%TCIMACRO{\tbigwedge \nolimits^{\diamond}}%
%BeginExpansion
{\textstyle\bigwedge\nolimits^{\diamond}}
%EndExpansion
V)$. Take $X\in%
%TCIMACRO{\tbigwedge \nolimits_{1}^{\diamond}}%
%BeginExpansion
{\textstyle\bigwedge\nolimits_{1}^{\diamond}}
%EndExpansion
V,$ $Y\in%
%TCIMACRO{\tbigwedge \nolimits_{3}^{\diamond}}%
%BeginExpansion
{\textstyle\bigwedge\nolimits_{3}^{\diamond}}
%EndExpansion
V$ and $\Phi\in%
%TCIMACRO{\tbigwedge \nolimits_{2}^{\diamond}}%
%BeginExpansion
{\textstyle\bigwedge\nolimits_{2}^{\diamond}}
%EndExpansion
V^{\ast},$ $\Psi\in%
%TCIMACRO{\tbigwedge \nolimits_{4}^{\diamond}}%
%BeginExpansion
{\textstyle\bigwedge\nolimits_{4}^{\diamond}}
%EndExpansion
V^{\ast}.$ A straightforward calculation, using Eq.(\ref{EPO18}),
Eq.(\ref{E7}), Eq.(\ref{EPO13}) and Eq.(\ref{EPO25}), gives
\begin{align*}
\underline{\lambda}\left\langle \tau,\sigma\right\vert (X,Y,\Phi,\Psi)  &
=\underline{\lambda}\circ\left\langle \tau,\sigma\right\vert (\underline
{\lambda}^{-1}(X),\underline{\lambda}^{-1}(Y),\underline{\lambda
}^{\bigtriangleup}(\Phi),\underline{\lambda}^{\bigtriangleup}(\Psi))\\
&  =\underline{\lambda}\left\langle \tau(\underline{\lambda}^{-1}%
(X),\underline{\lambda}^{\bigtriangleup}(\Phi)),\sigma(\underline{\lambda
}^{-1}(Y),\underline{\lambda}^{\bigtriangleup}(\Psi))\right\vert \\
&  =\left\langle \underline{\lambda}^{-\bigtriangleup}\circ\tau(\underline
{\lambda}^{-1}(X),\underline{\lambda}^{\bigtriangleup}(\Phi)),\underline
{\lambda}\circ\sigma(\underline{\lambda}^{-1}(Y),\underline{\lambda
}^{\bigtriangleup}(\Psi))\right\vert \\
&  =\left\langle \underline{\lambda}^{-\bigtriangleup}\tau(X,\Phi
),\underline{\lambda}\sigma(Y,\Psi)\right\vert \\
&  =\left\langle \underline{\lambda}^{-\bigtriangleup}\tau,\underline{\lambda
}\sigma\right\vert (X,Y,\Phi,\Psi),
\end{align*}
whence, the expected result follows.
\end{proof}

We present \ now some of the properties of the action of the extended of a
form operator $\lambda$ on multiform extensors.

\begin{itemize}
\item \textbf{ } For all $\tau\in\left.  \overset{\ast}{ext}\right.  _{k}%
^{l}(V)$ and $\sigma\in\left.  \overset{\ast}{ext}\right.  _{r}^{s}(V):$%
\begin{equation}
\underline{\lambda}(\tau\wedge\sigma)=(\underline{\lambda}\tau)\wedge
\underline{\lambda}(\sigma). \label{EPO28}%
\end{equation}

\item For all $\tau\in\left.  \overset{\ast}{ext}\right.  _{k}^{l}(V)$ and
$\sigma\in\left.  \overset{\left.  {}\right.  }{ext}\right.  _{r}^{s}(V):$%
\begin{align}
\underline{\lambda}\left\langle \tau,\sigma\right\rangle  &  =\left\langle
\underline{\lambda}\tau,\underline{\lambda}^{-\bigtriangleup}\sigma
\right\rangle ,\label{EPO29}\\
\underline{\lambda}\left\langle \tau,\sigma\right\vert  &  =\left\langle
\underline{\lambda}\tau,\underline{\lambda}^{-\bigtriangleup}\sigma\right\vert
,\label{EPO30}\\
\underline{\lambda}\left\vert \sigma,\tau\right\rangle  &  =\left\vert
\underline{\lambda}^{-\bigtriangleup}\sigma,\underline{\lambda}\tau
\right\rangle . \label{EPO31}%
\end{align}

\end{itemize}

\section{Generalization Procedure for Operators}

Let $\gamma$ be a linear operator on $V,$ i.e., a linear map $V\ni
v\longmapsto\gamma(v)\in V.$ It can be generalized in such a way to give a
linear operator on $%
%TCIMACRO{\tbigwedge }%
%BeginExpansion
{\textstyle\bigwedge}
%EndExpansion
V,$ namely $\underset{\smile}{\gamma},$ which is defined by
\[%
%TCIMACRO{\tbigwedge }%
%BeginExpansion
{\textstyle\bigwedge}
%EndExpansion
V\ni X\longmapsto\underset{\smile}{\gamma}(X)\in%
%TCIMACRO{\tbigwedge }%
%BeginExpansion
{\textstyle\bigwedge}
%EndExpansion
V
\]
such that%
\begin{equation}
\underset{\smile}{\gamma}(X)=\gamma(e_{j})\wedge\left\langle \varepsilon
^{j},X\right\vert , \label{GPO1}%
\end{equation}
where $\left\{  e_{j},\varepsilon^{j}\right\}  $ is any pair of dual bases for
$V$ and $V^{\ast}$.

We note that the multivector $\gamma(e_{j})\wedge\left\langle \varepsilon
^{j},X\right\vert $ does not depend on the choice of $\left\{  e_{j}%
,\varepsilon^{j}\right\}  ,$ and that the linearity of the duality contracted
product implies the linearity of $\underset{\smile}{\gamma}.$ Thus,
$\underset{\smile}{\gamma}$ is a well-defined linear operator on $%
%TCIMACRO{\tbigwedge }%
%BeginExpansion
{\textstyle\bigwedge}
%EndExpansion
V.$ We call $\underset{\smile}{\gamma}$ the \emph{generalized of }$\gamma$ (to
multivector operator).

The generalized of a vector operator $\gamma$ has the following basic properties.

\begin{itemize}
\item \textbf{ } $\underset{\smile}{\gamma}$ is grade-preserving, i.e.,
\begin{equation}
\text{if }X\in%
%TCIMACRO{\tbigwedge \nolimits^{k}}%
%BeginExpansion
{\textstyle\bigwedge\nolimits^{k}}
%EndExpansion
V,\text{ then }\underset{\smile}{\gamma}(X)\in%
%TCIMACRO{\tbigwedge \nolimits^{k}}%
%BeginExpansion
{\textstyle\bigwedge\nolimits^{k}}
%EndExpansion
V. \label{GPO2}%
\end{equation}

\item For all $\alpha\in\mathbb{R},$ $v\in V,$ and $X,Y\in%
%TCIMACRO{\tbigwedge }%
%BeginExpansion
{\textstyle\bigwedge}
%EndExpansion
V:$%
\begin{align}
\underset{\smile}{\gamma}(\alpha)  &  =0,\label{GPO3}\\
\underset{\smile}{\gamma}(v)  &  =\gamma(v),\label{GPO4}\\
\underset{\smile}{\gamma}(X\wedge Y)  &  =\underset{\smile}{\gamma}(X)\wedge
Y+X\wedge\underset{\smile}{\gamma}(Y). \label{GPO5}%
\end{align}

\end{itemize}

The four properties given by Eq.(\ref{GPO2}), Eq.(\ref{GPO3}). Eq.(\ref{GPO4})
and Eq.(\ref{GPO5}) are completely equivalent to the generalization procedure
for vector operators.

Let $\gamma$ be a linear operator on $V^{\ast}$, i.e., a linear map $V^{\ast
}\ni\omega\longmapsto\gamma(\omega)\in V^{\ast}$. It is possible to generalize
$\gamma$ in such a way as to get a linear operator on $%
%TCIMACRO{\tbigwedge }%
%BeginExpansion
{\textstyle\bigwedge}
%EndExpansion
V^{\ast},$ namely $\underset{\smile}{\gamma},$ which is defined by%
\[%
%TCIMACRO{\tbigwedge }%
%BeginExpansion
{\textstyle\bigwedge}
%EndExpansion
V^{\ast}\ni\Phi\longmapsto\underset{\smile}{\gamma}(\Phi)\in%
%TCIMACRO{\tbigwedge }%
%BeginExpansion
{\textstyle\bigwedge}
%EndExpansion
V^{\ast}%
\]
such that%
\begin{equation}
\underset{\smile}{\gamma}(\Phi)=\gamma(\varepsilon^{j})\wedge\left\langle
e_{j},\Phi\right\vert , \label{GPO6}%
\end{equation}
where $\left\{  e_{j},\varepsilon^{j}\right\}  $ is any pair of dual bases
over $V.$

We emphasize that $\underset{\smile}{\gamma}$ is a well-defined linear
operator on $%
%TCIMACRO{\tbigwedge }%
%BeginExpansion
{\textstyle\bigwedge}
%EndExpansion
V^{\ast}$, and call it the \emph{generalized of} $\gamma$ (to a multiform operator).

The generalized of a form operator $\gamma$ has the following basic properties.

\begin{itemize}
\item \textbf{ }$\underset{\smile}{\gamma}$ is grade-preserving, i.e.,
\begin{equation}
\text{if }\Phi\in%
%TCIMACRO{\tbigwedge \nolimits^{k}}%
%BeginExpansion
{\textstyle\bigwedge\nolimits^{k}}
%EndExpansion
V^{\ast},\text{ then }\underset{\smile}{\gamma}(\Phi)\in%
%TCIMACRO{\tbigwedge \nolimits^{k}}%
%BeginExpansion
{\textstyle\bigwedge\nolimits^{k}}
%EndExpansion
V^{\ast}. \label{GPO7}%
\end{equation}

\item \textbf{ }For all $\alpha\in\mathbb{R},$ $\omega\in V^{\ast},$ and
$\Phi,\Psi\in%
%TCIMACRO{\tbigwedge }%
%BeginExpansion
{\textstyle\bigwedge}
%EndExpansion
V^{\ast}$ we have
\begin{align}
\underset{\smile}{\gamma}(\alpha)  &  =0,\label{GPO8}\\
\underset{\smile}{\gamma}(\omega)  &  =\gamma(\omega),\label{GPO9}\\
\underset{\smile}{\gamma}(\Phi\wedge\Psi)  &  =\underset{\smile}{\gamma}%
(\Phi)\wedge\Psi+\Phi\wedge\underset{\smile}{\gamma}(\Psi). \label{GPO10}%
\end{align}

\end{itemize}

The properties given by Eq.(\ref{GPO7}), Eq.(\ref{GPO8}), Eq.(\ref{GPO9}) and
Eq.(\ref{GPO10}) are logically equivalent to the generalization procedure for
\textit{form} operators.

There exists a relationship between the generalization procedure of a vector
operator and the generalization procedure of a form operator.

Let $\gamma$ a vector operator (or, a form operator). As we already know, the
duality adjoint of $\gamma$ is just a form operator (respectively, a vector
operator), and the duality adjoint of $\underset{\smile}{\gamma}$ is just a
multiform operator(respectively, a multivector operator). The duality adjoint
of the generalized of $\gamma$ is equal to the generalized of the duality
adjoint of $\gamma,$ i.e.,%
\begin{equation}
\left(  \underset{\smile}{\gamma}\right)  ^{\bigtriangleup}=\underset{\smile
}{\left(  \gamma^{\bigtriangleup}\right)  }. \label{GPO11}%
\end{equation}

It follows that is possible to use a more simple notation, namely
$\underset{\smile}{\gamma}^{\bigtriangleup}$ to mean either $\left(
\underset{\smile}{\gamma}\right)  ^{\bigtriangleup}$ or $\underset{\smile
}{\left(  \gamma^{\bigtriangleup}\right)  }$.

We give some of the main properties of the generalized of a vector operator
$\gamma.$

\begin{itemize}
\item \textbf{\ }For all $\Phi\in%
%TCIMACRO{\tbigwedge }%
%BeginExpansion
{\textstyle\bigwedge}
%EndExpansion
V^{\ast},$ and $X\in%
%TCIMACRO{\tbigwedge }%
%BeginExpansion
{\textstyle\bigwedge}
%EndExpansion
V:$%
\begin{align}
\underset{\smile}{\gamma}\left\langle \Phi,X\right\rangle  &  =-\left\langle
\underset{\smile}{\gamma}^{\bigtriangleup}(\Phi),X\right\rangle +\left\langle
\Phi,\underset{\smile}{\gamma}(X)\right\rangle ,\label{GPO12}\\
\underset{\smile}{\gamma}\left\langle \Phi,X\right\vert  &  =-\left\langle
\underset{\smile}{\gamma}^{\bigtriangleup}(\Phi),X\right\vert +\left\langle
\Phi,\underset{\smile}{\gamma}(X)\right\vert ,\label{GPO13}\\
\underset{\smile}{\gamma}\left\vert X,\Phi\right\rangle  &  =\left\vert
\underset{\smile}{\gamma}(X),\Phi\right\rangle -\left\vert X,\underset{\smile
}{\gamma}^{\bigtriangleup}(\Phi)\right\rangle . \label{GPO14}%
\end{align}

\end{itemize}

We prove the property given by Eq.(\ref{GPO13}), the other proofs are analogous.

\begin{proof}
Let us take $X\in%
%TCIMACRO{\tbigwedge }%
%BeginExpansion
{\textstyle\bigwedge}
%EndExpansion
V$ and $\Phi,\Psi\in%
%TCIMACRO{\tbigwedge }%
%BeginExpansion
{\textstyle\bigwedge}
%EndExpansion
V^{\ast}.$ A straightforward calculation, by using Eq.(\ref{DAE6}),
Eq.(\ref{DCP8}), Eq.(\ref{GPO7}) and Eq.(\ref{GPO10}), yields%
\begin{align*}
\left\langle \underset{\smile}{\gamma}\left\langle \Phi,X\right\vert
,\Psi\right\rangle  &  =\left\langle \left\langle \Phi,X\right\vert
,\underset{\smile}{\gamma}^{\bigtriangleup}(\Psi)\right\rangle =\left\langle
X,\widetilde{\Phi}\wedge\underset{\smile}{\gamma}^{\bigtriangleup}%
(\Psi)\right\rangle \\
&  =\left\langle X,-\widetilde{\underset{\smile}{\gamma}^{\bigtriangleup}%
(\Phi)}\wedge\Psi+\underset{\smile}{\gamma}^{\bigtriangleup}(\widetilde{\Phi
})\wedge\Psi+\widetilde{\Phi}\wedge\underset{\smile}{\gamma}^{\bigtriangleup
}(\Psi)\right\rangle \\
&  =-\left\langle \left\langle \underset{\smile}{\gamma}^{\bigtriangleup}%
(\Phi),X\right\vert ,\Psi\right\rangle +\left\langle X,\underset{\smile
}{\gamma}^{\bigtriangleup}(\widetilde{\Phi}\wedge\Psi)\right\rangle \\
&  =-\left\langle \left\langle \underset{\smile}{\gamma}^{\bigtriangleup}%
(\Phi),X\right\vert ,\Psi\right\rangle +\left\langle \left\langle
\Phi,\underset{\smile}{\gamma}(X)\right\vert ,\Psi\right\rangle \\
&  =\left\langle -\left\langle \underset{\smile}{\gamma}^{\bigtriangleup}%
(\Phi),X\right\vert +\left\langle \Phi,\underset{\smile}{\gamma}(X)\right\vert
,\Psi\right\rangle ,
\end{align*}
and by the non-degeneracy of duality scalar product, the expected result follows.
\end{proof}

We present some properties for the generalized of a form operator $\gamma.$

\begin{itemize}
\item For all $\Phi\in%
%TCIMACRO{\tbigwedge }%
%BeginExpansion
{\textstyle\bigwedge}
%EndExpansion
V^{\ast},$ and $X\in%
%TCIMACRO{\tbigwedge }%
%BeginExpansion
{\textstyle\bigwedge}
%EndExpansion
V:$%
\begin{align}
\underset{\smile}{\gamma}\left\langle \Phi,X\right\rangle  &  =\left\langle
\underset{\smile}{\gamma}(\Phi),X\right\rangle -\left\langle \Phi
,\underset{\smile}{\gamma}^{\bigtriangleup}(X)\right\rangle ,\label{GPO15}\\
\underset{\smile}{\gamma}\left\langle \Phi,X\right\vert  &  =\left\langle
\underset{\smile}{\gamma}(\Phi),X\right\vert -\left\langle \Phi,\underset
{\smile}{\gamma}^{\bigtriangleup}(X)\right\vert ,\label{GPO16}\\
\underset{\smile}{\gamma}\left\vert X,\Phi\right\rangle  &  =-\left\vert
\underset{\smile}{\gamma}^{\bigtriangleup}(X),\Phi\right\rangle +\left\vert
X,\underset{\smile}{\gamma}(\Phi)\right\rangle . \label{GPO17}%
\end{align}

\end{itemize}

\subsection{Action of Generalized Operators on Extensors}

Let $\gamma$ be a linear operator on $V$. We can generalize the action of
$\underset{\smile}{\gamma}$ in such a way $\underset{\smile}{\gamma}$ is to
map multivector extensors over $V$ into multivector extensors over $V$. We
define the linear\ mapping%
\[
\left.  \overset{\left.  {}\right.  }{ext}\right.  _{k}^{l}(V)\ni
\tau\longmapsto\underset{\smile}{\gamma}\tau\in\left.  \overset{\left.
{}\right.  }{ext}\right.  _{k}^{l}(V)
\]
such that%
\begin{align}
\underset{\smile}{\gamma}\tau(X_{1},\ldots,X_{k},\Phi^{1},\ldots,\Phi^{l})  &
=\underset{\smile}{\gamma}\circ\tau(X_{1},\ldots,X_{k},\Phi^{1},\ldots
,\Phi^{l})\nonumber\\
&  -\tau(\underset{\smile}{\gamma}(X_{1}),\ldots,X_{k},\Phi^{1},\ldots
,\Phi^{l})\nonumber\\
&  \cdots-\tau(X_{1},\ldots,\underset{\smile}{\gamma}(X_{k}),\Phi^{1}%
,\ldots,\Phi^{l})\nonumber\\
&  +\tau(X_{1},\ldots,X_{k},\underset{\smile}{\gamma}^{\bigtriangleup}%
(\Phi^{1}),\ldots,\Phi^{l})\nonumber\\
&  \cdots+\tau(X_{1},\ldots,X_{k},\Phi^{1},\ldots,\underset{\smile}{\gamma
}^{\bigtriangleup}(\Phi^{l}) \label{GPO18}%
\end{align}
for each $X_{1},\ldots,X_{k}\in%
%TCIMACRO{\tbigwedge }%
%BeginExpansion
{\textstyle\bigwedge}
%EndExpansion
V$ and $\Phi^{1},\ldots,\Phi^{l}\in%
%TCIMACRO{\tbigwedge }%
%BeginExpansion
{\textstyle\bigwedge}
%EndExpansion
V^{\ast}.$

Thus, we can think of $\underset{\smile}{\gamma}$ as a linear multivector
extensor operator.

For instance, for $\tau\in\left.  \overset{\left.  {}\right.  }{ext}\right.
_{1}^{0}(V)$ this definition above gives%
\begin{equation}
\underset{\smile}{\gamma}\tau=\underset{\smile}{\gamma}\circ\tau-\tau
\circ\underset{\smile}{\gamma}=\left[  \underset{\smile}{\gamma},\tau\right]
, \label{GPO19}%
\end{equation}
and for $\tau\in\left.  \overset{\left.  {}\right.  }{ext}\right.  _{0}%
^{1}(V)$ it yields%
\begin{equation}
\underset{\smile}{\gamma}\tau=\underset{\smile}{\gamma}\circ\tau+\tau
\circ\underset{\smile}{\gamma}^{\bigtriangleup}. \label{GPO20}%
\end{equation}

Let $\gamma$ be an invertible linear operator on $V^{\ast}.$ Analogously to
the case above, it is possible to generalize the action of $\underset{\smile
}{\gamma}$ in such a way to get a linear operator on $\left.  \overset{\left.
\ast\right.  }{ext}\right.  _{k}^{l}(V)$. We define%
\[
\left.  \overset{\left.  \ast\right.  }{ext}\right.  _{k}^{l}(V)\ni
\tau\longmapsto\underset{\smile}{\gamma}\tau\in\left.  \overset{\left.
\ast\right.  }{ext}\right.  _{k}^{l}(V)
\]
such that%
\begin{align}
\underset{\smile}{\gamma}\tau(X_{1},\ldots,X_{k},\Phi^{1},\ldots,\Phi^{l})  &
=\underset{\smile}{\gamma}\circ\tau(X_{1},\ldots,X_{k},\Phi^{1},\ldots
,\Phi^{l})\nonumber\\
&  +\tau(\underset{\smile}{\gamma}^{\bigtriangleup}(X_{1}),\ldots,X_{k}%
,\Phi^{1},\ldots,\Phi^{l})\nonumber\\
&  \cdots+\tau(X_{1},\ldots,\underset{\smile}{\gamma}^{\bigtriangleup}%
(X_{k}),\Phi^{1},\ldots,\Phi^{l})\nonumber\\
&  -\tau(X_{1},\ldots,X_{k},\underset{\smile}{\gamma}(\Phi^{1}),\ldots
,\Phi^{l})\nonumber\\
&  \cdots-\tau(X_{1},\ldots,X_{k},\Phi^{1},\ldots,\underset{\smile}{\gamma
}(\Phi^{l}) \label{GPO25}%
\end{align}
for each $X_{1},\ldots,X_{k}\in%
%TCIMACRO{\tbigwedge }%
%BeginExpansion
{\textstyle\bigwedge}
%EndExpansion
V$ and $\Phi^{1},\ldots,\Phi^{l}\in%
%TCIMACRO{\tbigwedge }%
%BeginExpansion
{\textstyle\bigwedge}
%EndExpansion
V^{\ast}.$

For instance, for $\tau\in\left.  \overset{\ast}{ext}\right.  _{1}^{0}(V)$ the
definition just given above yields%
\begin{equation}
\underset{\smile}{\gamma}\tau=\underset{\smile}{\gamma}\circ\tau+\tau
\circ\underset{\smile}{\gamma}^{\bigtriangleup}, \label{GPO26}%
\end{equation}
and for $\tau\in\left.  \overset{\left.  \ast\right.  }{ext}\right.  _{0}%
^{1}(V)$ it holds%
\begin{equation}
\underset{\smile}{\gamma}\tau=\underset{\smile}{\gamma}\circ\tau-\tau
\circ\underset{\smile}{\gamma}=\left[  \underset{\smile}{\gamma},\tau\right]
. \label{GPO27}%
\end{equation}

We present some of the main properties of the action of the generalized
operator of a vector operator $\gamma$ on multivector extensors.

\begin{itemize}
\item For all $\tau\in\left.  \overset{\left.  {}\right.  }{ext}\right.
_{k}^{l}(V)$ and $\sigma\in\left.  \overset{\left.  {}\right.  }{ext}\right.
_{r}^{s}(V):$%
\begin{equation}
\underset{\smile}{\gamma}(\tau\wedge\sigma)=(\underset{\smile}{\gamma}%
\tau)\wedge\sigma+\tau\wedge(\underset{\smile}{\gamma}\sigma). \label{GPO21}%
\end{equation}

\item For all $\tau\in\left.  \overset{\ast}{ext}\right.  _{k}^{l}(V)$ and
$\sigma\in\left.  \overset{\left.  {}\right.  }{ext}\right.  _{r}^{s}(V):$%
\begin{align}
\underset{\smile}{\gamma}\left\langle \tau,\sigma\right\rangle  &
=-\left\langle \underset{\smile}{\gamma}^{\bigtriangleup}\tau,\sigma
\right\rangle +\left\langle \tau,\underset{\smile}{\gamma}\sigma\right\rangle
,\label{GPO22}\\
\underset{\smile}{\gamma}\left\langle \tau,\sigma\right\vert  &
=-\left\langle \underset{\smile}{\gamma}^{\bigtriangleup}\tau,\sigma
\right\vert +\left\langle \tau,\underset{\smile}{\gamma}\sigma\right\vert
,\label{GPO23}\\
\underset{\smile}{\gamma}\left\vert \sigma,\tau\right\rangle  &  =\left\vert
\underset{\smile}{\gamma}\sigma,\tau\right\rangle -\left\vert \sigma
,\underset{\smile}{\gamma}^{\bigtriangleup}\tau\right\rangle . \label{GPO24}%
\end{align}

\end{itemize}

\begin{proof}
We present only the proof for the property given by Eq. (\ref{GPO21}), the
others are similar. Without any loss of generality, we give the proof for the
particular case where $\tau\in ext(%
%TCIMACRO{\tbigwedge \nolimits_{1}^{\diamond}}%
%BeginExpansion
{\textstyle\bigwedge\nolimits_{1}^{\diamond}}
%EndExpansion
V,%
%TCIMACRO{\tbigwedge \nolimits_{2}^{\diamond}}%
%BeginExpansion
{\textstyle\bigwedge\nolimits_{2}^{\diamond}}
%EndExpansion
V^{\ast};%
%TCIMACRO{\tbigwedge \nolimits^{\diamond}}%
%BeginExpansion
{\textstyle\bigwedge\nolimits^{\diamond}}
%EndExpansion
V^{\ast})$ and $\sigma\in ext(%
%TCIMACRO{\tbigwedge \nolimits_{3}^{\diamond}}%
%BeginExpansion
{\textstyle\bigwedge\nolimits_{3}^{\diamond}}
%EndExpansion
V,%
%TCIMACRO{\tbigwedge \nolimits_{4}^{\diamond}}%
%BeginExpansion
{\textstyle\bigwedge\nolimits_{4}^{\diamond}}
%EndExpansion
V^{\ast};%
%TCIMACRO{\tbigwedge \nolimits^{\diamond}}%
%BeginExpansion
{\textstyle\bigwedge\nolimits^{\diamond}}
%EndExpansion
V)$. Take $X\in%
%TCIMACRO{\tbigwedge \nolimits_{1}^{\diamond}}%
%BeginExpansion
{\textstyle\bigwedge\nolimits_{1}^{\diamond}}
%EndExpansion
V,$ $Y\in%
%TCIMACRO{\tbigwedge \nolimits_{3}^{\diamond}}%
%BeginExpansion
{\textstyle\bigwedge\nolimits_{3}^{\diamond}}
%EndExpansion
V$ and $\Phi\in%
%TCIMACRO{\tbigwedge \nolimits_{2}^{\diamond}}%
%BeginExpansion
{\textstyle\bigwedge\nolimits_{2}^{\diamond}}
%EndExpansion
V^{\ast},$ $\Psi\in%
%TCIMACRO{\tbigwedge \nolimits_{4}^{\diamond}}%
%BeginExpansion
{\textstyle\bigwedge\nolimits_{4}^{\diamond}}
%EndExpansion
V^{\ast}.$ By using the definition of $\underset{\smile}{\gamma}$\ , we have%
\begin{equation}%
\begin{array}
[c]{ll}%
\underset{\smile}{\gamma}(\tau\wedge\sigma)\left(  X,Y,\Phi,\Psi\right)  &
=\underset{\smile}{\gamma}\circ(\tau\wedge\sigma)\left(  X,Y,\Phi,\Psi\right)
-(\tau\wedge\sigma)\left(  \underset{\smile}{\gamma}X,Y,\Phi,\Psi\right) \\
& -(\tau\wedge\sigma)\left(  X,\underset{\smile}{\gamma}Y,\Phi,\Psi\right)
+(\tau\wedge\sigma)\left(  X,Y,\underset{\smile}{\gamma}\Phi,\Psi\right) \\
& +(\tau\wedge\sigma)\left(  X,Y,\Phi,\underset{\smile}{\gamma}\Psi\right)  .
\end{array}
\label{EN4}%
\end{equation}
Now, using the property (\ref{GPO5}) we can write the first term of right side
of the Eq. (\ref{EN4}) as%
\[
\underset{\smile}{\gamma}\circ(\tau\wedge\sigma)\left(  X,Y,\Phi,\Psi\right)
=\underset{\smile}{\gamma}\circ\tau\left(  X,\Phi\right)  \wedge\sigma\left(
Y,\Psi\right)  +\tau\left(  X,\Phi\right)  \wedge\underset{\smile}{\gamma
}\circ\sigma\left(  Y,\Psi\right)  ,
\]
and remembering that
\[
(\tau\wedge\sigma)\left(  X,Y,\Phi,\Psi\right)  =\tau\left(  X,\Phi\right)
\wedge\sigma\left(  Y,\Psi\right)  ,
\]
the Eq. (\ref{EN4}) can be written as
\[%
\begin{array}
[c]{ll}%
\underset{\smile}{\gamma}(\tau\wedge\sigma)\left(  X,Y,\Phi,\Psi\right)  &
=\left[  \underset{\smile}{\gamma}\circ\tau\left(  X,\Phi\right)  -\tau\left(
\underset{\smile}{\gamma}X,\Phi\right)  +\tau\left(  X,\underset{\smile
}{\gamma^{\bigtriangleup}}\Phi\right)  \right]  \wedge\sigma\left(
Y,\Psi\right) \\
& +\tau\left(  X,\Phi\right)  \wedge\left[  \underset{\smile}{\gamma}%
\circ\sigma\left(  Y,\Psi\right)  -\sigma\left(  \underset{\smile}{\gamma
}Y,\Psi\right)  +\sigma\left(  Y,\underset{\smile}{\gamma^{\bigtriangleup}%
}\Psi\right)  \right]
\end{array}
\]
or
\[%
\begin{array}
[c]{ll}%
\underset{\smile}{\gamma}(\tau\wedge\sigma)\left(  X,Y,\Phi,\Psi\right)  &
=\left[  \left(  \underset{\smile}{\gamma}\tau\right)  \left(  X,\Phi\right)
\right]  \wedge\sigma\left(  Y,\Psi\right)  +\tau\left(  X,\Phi\right)
\wedge\left[  \left(  \underset{\smile}{\gamma}\sigma\right)  \left(
Y,\Psi\right)  \right] \\
& =\left(  \underset{\smile}{\gamma}\tau\wedge\sigma+\tau\wedge\underset
{\smile}{\gamma}\sigma\right)  \left(  X,Y,\Phi,\Psi\right)  ,
\end{array}
\]
and the property is proved.
\end{proof}

We present some properties for the action of the generalized operator of a
form operator $\gamma$ on multiform extensors.

\begin{itemize}
\item For all $\tau\in\left.  \overset{\left.  \ast\right.  }{ext}\right.
_{k}^{l}(V)$ and $\sigma\in\left.  \overset{\left.  \ast\right.  }%
{ext}\right.  _{r}^{s}(V):$%
\begin{equation}
\underset{\smile}{\gamma}(\tau\wedge\sigma)=(\underset{\smile}{\gamma}%
\tau)\wedge\sigma+\tau\wedge(\underset{\smile}{\gamma}\sigma). \label{GPO28}%
\end{equation}

\item \textbf{ } For all $\tau\in\left.  \overset{\ast}{ext}\right.  _{k}%
^{l}(V)$ and $\sigma\in\left.  \overset{\left.  {}\right.  }{ext}\right.
_{r}^{s}(V):$%
\begin{align}
\underset{\smile}{\gamma}\left\langle \tau,\sigma\right\rangle  &
=\left\langle \underset{\smile}{\gamma}\tau,\sigma\right\rangle -\left\langle
\tau,\underset{\smile}{\gamma}^{\bigtriangleup}\sigma\right\rangle
,\label{GPO29}\\
\underset{\smile}{\gamma}\left\langle \tau,\sigma\right\vert  &  =\left\langle
\underset{\smile}{\gamma}\tau,\sigma\right\vert -\left\langle \tau
,\underset{\smile}{\gamma}^{\bigtriangleup}\sigma\right\vert ,\label{GPO30}\\
\underset{\smile}{\gamma}\left\vert \sigma,\tau\right\rangle  &  =-\left\vert
\underset{\smile}{\gamma}^{\bigtriangleup}\sigma,\tau\right\rangle +\left\vert
\sigma,\underset{\smile}{\gamma}\tau\right\rangle . \label{GPO31}%
\end{align}

\end{itemize}

\section{Conclusions}

In this paper we studied the properties of the \textit{duality product} of
multivectors and multiforms (used for the definition of the hyperbolic
Clifford algebra of \textit{multivefors \cite{qr07}}) \ introducing a very
useful notation for the left and right contracted products of multiforms
(elements of $\bigwedge V^{\ast}$) and multivectors (elements of $\bigwedge
V$). Next, we introduced a theory of the $k$\emph{\ multivector and }%
$l$\emph{\ multiform variables multivector }(\emph{or multiform}%
)\emph{\ extensors} over $V$ (defining the spaces $\left.  \overset{\left.
{}\right.  }{ext}\right.  _{k}^{l}(V)$ and $\left.  \overset{\left.
\ast\right.  }{ext}\right.  _{k}^{l}(V)$) defining the exterior product of
extensors, and several important operations on them. This algebraic theory
will play an important role in a presentation of the differential geometry of
a manifold $M$ of arbitrary topology discussed in forthcoming papers.

\end{document}